# Coherence, long-range transport and nuclear polarization in a driven-dissipative dark exciton condensate

Amit Jash, Maheswar Swar, Uri Shimon, Vladimir Umansky, and Israel Bar-Joseph

*Department of Condensed Matter physics, Weizmann Institute of Science, Rehovot 7610001, Israel*

We report direct evidence for macroscopic coherence in a condensate of dark dipolar excitons in coupled quantum wells and show that its formation follows a non-equilibrium, driven-dissipative mechanism. The condensation transition is governed by gain-loss competition, in which the exceptionally long lifetime of dark excitons enables their dominance in mode selection. Condensate formation is revealed by photoluminescence darkening, changes in radiative recombination channels, and the emergence of long-range hydrodynamic transport manifested by propagation of density (sound) modes over millimeter-scale distances. The buildup of dark exciton density induces dynamic nuclear polarization, which closes the dark-bright exciton gap, $\Delta$, via the Overhauser field. This leads to nuclear spin polarization across the entire mesa, far beyond the optically excited region, and produces pronounced hysteresis behavior. At $\Delta \approx 0$ the gap is locked and the condensate loss are minimal, resulting in a second threshold manifested as a photoluminescence blueshift. Coherence is revealed through interference between incident and boundary-reflected exciton currents, producing spatial modulation of the photoluminescence from the radiative reservoir and enabling extraction of the condensate coherence length. These results establish dark excitons as a platform for coherent quantum fluids in a driven-dissipative, strongly interacting regime with electrical tunability, bridging the physics of polariton condensates and matter-like excitonic systems.

## Introduction

Dipolar indirect excitons (IX) in coupled quantum wells (CQWs) have been long considered as attractive candidates for realizing equilibrium condensate [1-20]. Dark IX, whose radiative recombination is forbidden by spin selection rules, have attracted particular attention in this context [11-20]. Their suppressed radiative decay allows densities and lifetimes far exceeding those of bright excitons or polaritons. Unfortunately, however, this same darkness complicates experimental access and obscures the nature of their condensation. Indeed, previous studies have inferred condensation indirectly, primarily from interaction-induced photoluminescence (PL) blueshift, which is observed below a critical temperature and above a threshold optical power in CQWs, and is interpreted as an increase in the dark exciton density [14,16-19]. In this picture, condensation is commonly interpreted within an equilibrium Bose-Einstein condensation (BEC) framework, where the bright exciton band acts as a reservoir that feeds the lower energy dark band [17].

Several conceptual and experimental inconsistencies have accumulated over time with the equilibrium BEC picture of dark exciton condensation. First, in homogeneous two-dimensional systems, conventional equilibrium BEC is forbidden for an ideal Bose gas. A common route around this restriction is harmonic confinement, which discretizes the spectrum and modifies the density of states, enabling macroscopic ground-state occupation. Indeed, many experiments intentionally employ electrostatic or strain-induced traps to facilitate condensation [6,15,16,21]. However, threshold blueshift in power and temperature has also been observed in open geometry CQWs experiments without intentional confinement [8,9,17]. In such systems, a thermodynamic interpretation is difficult to justify. An even more striking inconsistency arises in the presence of a magnetic field applied parallel to the CQWs layers. When the electron Zeeman energy becomes comparable to the bright-dark exciton gap, Δ, strong mixing between dark and bright states is expected. The ground state should then acquire radiative character, and accumulation of a long-lived dark excitons population would be incompatible with an equilibrium description. Experimentally, however, no change in the behavior is observed even when the applied field exceeds the nominal critical value by two orders of magnitude [19,20]. These observations call for a reexamination of the equilibrium BEC paradigm for dark excitons.

In this work, we provide direct evidence for coherence in a dark exciton condensate and demonstrate that its formation follows a non-equilibrium route. Rather than emerging from thermodynamic equilibration, condensation is governed by gain-loss competition in a driven-dissipative system. Owing to their exceptionally low decay rate, dark excitons dominate the mode-selection process, analogous to lasing. This mechanism closely parallels condensation in microcavity polaritons [22], with the crucial distinction that the order parameter here is optically dark. We show that condensation in this system is marked by a qualitative change in transport: below threshold, excitons diffuse over short distances, whereas above threshold they form a hydrodynamic quantum fluid, allowing the condensate to extend across the entire mesa and support collective density (sound) waves with well-defined velocity. When the system is excited near the mesa boundary, the interference between the outgoing condensate wavefunction and the reflected one from the boundary yields spatial modulation of the PL emission, providing direct evidence for condensate coherence. We demonstrate a unique property of the condensate that arises from its interaction with nuclei, dynamic polarization of the nuclei by the condensate, which closes the bright-dark exciton gap, $\Delta$, and gives rise to hysteresis behavior in power and magnetic field. Employing RF spectroscopy, we show that above a critical power the gap is locked at $\Delta \approx 0$, yielding dark exciton density buildup.

**Evidence for dark IX condensation via driven-dissipative dynamics**

The non-equilibrium route to dark exciton condensation can be understood by considering exciton formation and relaxation in CQWs. Consider a CQWs consisting of wide (WW) and narrow (NW) wells, separated by a thin barrier and subjected to an electric field that sets the electron level in the NW lower than the WW (Fig. 1a and see section 1 of the supplementary material (SM)). Electrons that are photoexcited in the WW may tunnel and create hot population with in-plane momentum $\boldsymbol{k}_e$ and distribution $f(\boldsymbol{k}_e)$ in the NW (Fig. 1b). These electrons cool down by emitting phonons with momentum $\boldsymbol{q}$ and are captured by the cold holes left behind in the WW at a rate $R(\boldsymbol{q})$, forming IX with center of mass momentum $\boldsymbol{K} = \boldsymbol{k}_e - \boldsymbol{q}_{\parallel}$, where $\boldsymbol{q}_{\parallel}$ is the in-plane phonon momentum. The rate of this capture process, $G(\boldsymbol{K}) \propto \int d^2\boldsymbol{k}_e\, f_e(\boldsymbol{k}_e) R(|\boldsymbol{k}_e - \boldsymbol{K}|)$, is a decreasing function of $|\boldsymbol{K}|$ [see section 2 of the SM], hence, after a multi-phonon relaxation and exciton-exciton scattering, the IX population forms a Bose distribution that peaks at $K \approx 0$. Once a small seed at this state exists with occupation $N_0$, Bose stimulation multiplies its feed by $(1 + N_0)$, and if the gain of this mode, $n_R G(0)$, exceeds its loss, $\gamma_0$, it wins the mode competition, similarly to laser threshold. Here, $n_R$ is the electron-hole reservoir density and $\gamma_0$ is the loss rate of the $K \approx 0$ state. Crucially, because dark excitons have loss rates orders of magnitude smaller than those of bright excitons, they reach threshold at much lower excitation power (Fig. 1c, left panel). *Thus, it is their exceptionally low dissipation rather than their lower energy that allows dark excitons to win the mode competition.*

***Condensation threshold*:** To probe the condensation process experimentally, we exploit the fact that a fraction of photoexcited electrons in the WW remain in that well and form positively charged trions, T, with the abundant holes (Fig. 1a). Since T formation is an intra-well process, it occurs on a faster time scale than the competing inter-well IX forming process. Indeed, the T emission line dominates the spectrum at low power (Fig. 1d). However, its binding energy (~1 meV) is smaller than that of the IX (~2.5 meV), and it may dissociate through capturing a NW electron and emitting a phonon into indirect and direct (DX) excitons, $\mathrm{T} + \mathrm{e} \rightarrow \mathrm{IX} + \mathrm{DX}$. This process is Bose-stimulated when a dark IX population builds up, giving rise to diminishing trion PL intensity, $I_T$, and increasing DX intensity, $I_{DX}$, thereby providing a direct sensitive probe of condensation. Taking the ratio between the trion dissociation and recombination rates to be $R_T$ and the excitation power as $P$, we can express $I_T$ as $\propto P/[1 + R_T(1 + N_0)] \approx P[1 - R_T N_0]$. Hence, we expect the normalized trion intensity, $\bar{I}_T = I_T/P$, to be constant at $R_T N_0 \ll 1$, and to drop above threshold at a slope that is proportional to the ground state occupation $N_0$.

The measured power dependence of the PL spectrum indeed exhibits this behavior. We find that $\bar{I}_T$ is constant up to $P_{th} \approx 100$ nW and then drops at higher power, while the normalized DX intensity, $\bar{I}_{DX} = I_{DX}/P$, shows the opposite trend (Fig. 2a). Concurrently, the *total* normalized PL intensity emitted from the sample, $\bar{I}_{Tot} = I_{Tot}/P$, (where $I_{tot}$ is the sum of the T, DX, and bright IX lines), which is constant at $P < P_{th}$, falls above $P_{th}$, indicating the onset of non-radiative processes. The non-radiative recombination rate, $\propto \gamma_{NR} N_0$, can be obtained by subtracting $I_{Tot}$ from a linear interpolation of the fully radiative subthreshold behavior (Fig. 2b). Assuming $\gamma_{NR}$ is constant, we can extract the evolution of $N_0$ with power. It is seen that it grows with power by ~ two orders of magnitude above threshold at approximately a linear rate, reflecting the expected clamping of the reservoir density at the condensation onset. Remarkably, the PL darkening occurs uniformly across the pump beam area, confirming its global nature and ruling out defect related processes (See Extended Data, Figs. S7-8).

We find that $P_{th}$ is strongly affected by applying a magnetic field normal to the layers, decreasing $P_{th}$ from 100 nW at $B = 0$ down to a few nW at $B = 1.5$T. This further rule out extrinsic processes, such as screening of the disorder potential, as responsible for the change in behavior at $P_{th}$. Rather, the observed strong dependence on magnetic field indicates that the dominant loss mechanism is spin flip processes, whose rate scales as $1/\Delta^2$ (Fig. 1c, left panel), where $\Delta \approx \mathrm{g_e}\mu_B B$ is the electron Zeeman energy, and is quenched with increasing magnetic field (see Extended Data, Fig. S9).

Remarkably, the power values at which threshold occurs remain approximately the same while varying the spot size between 20 to 1 μm, corresponding to modifying the local power density by more than two orders of magnitude (Fig. 2c). This suggests that the excitons spread from the excitation region, such that the threshold density is defined over a constant large area rather than the excitation spot area. This expansion is confirmed by pump-probe experiments, where the spectrum at remote locations away from the pump is measured by a weak probe. Despite the large separation, the trion intensity of the probe exhibits a similar decrease above a threshold pump power (Fig. 2d), indicating that the pump threshold is sensed throughout the mesa.

***Density build-up and blueshift*:** A conclusive proof that dark excitons fill the entire area of the mesa is provided by PL measurements in a magnetic field. Figure 2e depicts the blueshift of the IX energy with pump power at the pump and remote probe locations. This blueshift, which is due to the repulsive interaction between the dipolar excitons, allows us to extract their density: the measured blueshift slope, $\sim 0.15$ meV/μW, corresponds to IX density slope of $2-3\times 10^9$ cm$^{-2}$/μW [5]. Remarkably, we find the *same* blueshift in both locations, implying that the *same* IX density builds up throughout the $500\times 100$ μm$^2$ mesa. Taking the absorbed power to be $\sim 2\%$ of the incident power, we can estimate the exciton lifetime to be $10-20$ μs. These values are ~ two orders of magnitude larger than the bright excitons lifetime, which is measured to be 70 ns at this gate voltage, and are consistent with dark exciton lifetime. The emerging picture is then of a condensate of dark excitons that is continuously fed by the localized optical pump and spreads away from the excitation region, establishing a constant steady-state density in the entire mesa within the dark exciton's lifetime.

The linear dependence of the dark exciton density on power (Figs. 1d and 2e) prevails up to a critical power [20], where we observe a second threshold: The IX energy sharply blueshifts [14,16-19] and the T line vanishes (Fig. 1d), reflecting a nonlinear density increase with increasing power. This second threshold is observed throughout the mesa [20], indicating that it reflects a global change in the system properties. Time-resolved measurements reveal a fast response of the probe spectrum when the pump power is modulated by a square wave, with a characteristic time delay of tens of ns that depends on the probe position (Fig. 2f). Because the modulation period, 2 μs, is shorter than the dark exciton lifetime, the pump modulation generates only a small density perturbation on the exciton fluid, which propagates as a density wave throughout the mesa. The measured velocity of $\sim 2 \times 10^4$ m/s matches the expected sound velocity of a dipolar exciton liquid for a density of $\sim 10^{10}$ cm$^{-2}$, indicating collective dynamics [23] rather than single-particle diffusion. We note that at this velocity, density waves can propagate over tens of centimeters within the dark exciton lifetime, thereby establishing a global steady state upon reflection from the boundaries.

**Evidence for dark condensate coherence**

Writing the wave function in a constant density condensate as $\psi = \sqrt{n}\exp[i\Phi(x,y)]$, where $\Phi(x,y)$ is the condensate phase field, the steady-state exciton current away from the pump beam is $J = n\frac{\hbar}{m}\nabla\Phi$. Near a boundary the phase gradient should give rise to an interference density pattern $\propto 2r\cos(2kx)$ between incoming, $\exp(ikx)$, and reflected, $r\exp(-ikx)$, exciton currents, where $k$ is defined as $k = \nabla_{\rm x}\Phi$ and $r$ is the reflection coefficient. This interference pattern is indeed observed when we position the pump beam close to the mesa boundary (inset of Fig. 3a): clear fringes, which are parallel to the boundary, appear in the PL image (Fig. 3a and Fig. S12 of the Extended Data). Using a small iris in the image plane and measuring the spectrum we verified that these fringes represent modulation of the PL emission intensity from the sample. We conclude that spatial modulation of the dark exciton density modulates the stimulated scattering rate from the reservoir. Hence, a high-density region would be manifested as low emission intensity and a low density – as high intensity. Indeed, the observed pattern exhibits a node at the boundary, and is proportional to $\sin(2kx)$.

Figure 3b depicts the fringe visibility curve, $[I(x) - \bar{I}]/\bar{I}$, where $I(x)$ and $\bar{I}$ are the local and mean PL values, respectively. We find that the fringe spatial frequency, $k(x)$, exhibits a chirped behavior, $k(x) = k_0 + \alpha x$, where $k_0 = 0.9\ \mu\text{m}^{-1}$ is the value of $k$ at the boundary and $\alpha \approx 0.1\ \mu\text{m}^{-2}$. This chirped behavior is expected for current flowing in a lossy medium: Assuming a constant density $n$ and loss $\gamma$, the continuity equation, $dJ/dx = \gamma n$, yields $k = k_0 + \frac{\gamma m}{\hbar} x$. Confirmation for this interpretation is obtained by measuring the interference pattern as we vary the distance of the pump beam center relative to the boundary, $L$ (Fig. 3c). We find that the interference pattern periodicity decreases with increasing $L$, such that $k_0$, extracted from the period of the first fringe near the boundary, drops linearly with $L$ at a slope $-\alpha L$ (Fig. 3d). This implies that the pump beam creates a global phase field, $\Phi$, with a constant gradient near the boundary, $\nabla_x \Phi = k$, and the condensate current flows along this phase gradient and is attenuated at a rate $\gamma n$.

The interference pattern envelope exhibits an exponential decay with distance from the boundary (Figs. 3b-c and Fig. S12 of the Extended Data). Since this pattern is proportional to the condensate first order correlation function, $g^{(1)}(x, -x) \propto \langle \psi^*(x\,)\psi(-x) \rangle$, we can extract the condensate coherence length, $l_\phi$, from the envelope decay, $\exp(-x/l)$. Note that in reflection geometry, $l_\phi$ is twice the decay length of the fringe envelope. Figure 3e depicts $l_\phi$ as a function of pump position, $L/\sigma$, where $\sigma$ is the Gaussian beam radius, demonstrating the increase of the coherence length as we move away from the beam center, approaching $l_\phi \sim 10\ \mu\text{m}$ at $\sim 3\sigma$.

The interference fringes are observed near the mesa boundaries that are parallel to the y-axis and are not observed near the orthogonal boundaries along x-axis. Scanning electron and atomic force microscopy imaging reveal that the y boundaries have a pyramid-shape, such that the CQWs extends $\sim 1\ \mu\text{m}$ beyond the top gate layer end [see Fig. S2 of the SM]. This creates a smooth and high electrostatic potential barrier for the dipolar excitons before the actual CQW edge, which reflects the excitons coherently. The x-boundary, on the other hand, is nearly vertical due to the non-isotropic etching. The edge of the CQW is defined in this case by the lossy etched surface and quenches the interference.

**Condensation and nuclear polarization**

A unique property of the dark condensate arises from its interaction with the nuclei. We have previously shown that the buildup of dark exciton density gives rise to polarization of the NW nuclear spins, such that the resulting Overhauser field closes the dark-bright exciton gap, $\Delta$ [20]. The underlying mechanism is hyperfine interaction mediated flip-flop transition, $\uparrow_e \downarrow_N \rightleftarrows \downarrow_e \uparrow_N$ [24-30], which turns a dark exciton with electron spin $\uparrow_e$ into bright exciton with electron spin $\downarrow_e$. The fast recombination of the bright exciton leaves the nuclear system in the $\uparrow_N$ polarization, thereby imprinting the spin orientation of the electron in the dark exciton in the nuclear polarization [20]. The rate equation governing the buildup of nuclear polarization $P_N \in (-1, +1)$ is,

$$dP_N/dt = (W_+ - W_-) - (W_+ + W_- + \Gamma_N)\, P_N, \quad (1)$$

where $\Gamma_N$ is the decay rate, and $W_\pm$ are the polarization and depolarization rates, both $\propto 1/(\Gamma^2 + \Delta^2)$, where $\Gamma$ is the width of the bright-dark resonance. The buildup of $P_N$ yields to an Overhauser shift, $AP_N$, that decreases the zero-power dark-bright gap, $\Delta_0$, [31] such that $\Delta = \Delta_0 - AP_N$. As a result, Eq. (1) is nonlinear in $P_N$, and exhibits a bistable behavior with two branches: one at which $P_N \approx 0$ and the other at $P_N \approx \Delta_0/A$ [29, see section 3 of the SM]. At low power, the condensate-nuclei system is at the low polarization branch, and the dark exciton band is decoupled from the bright band. However, above a certain power, the system switches into the high polarization branch, where $\Delta \approx 0$ (Fig. 1c, right panel). As the pump power is further increased, the system locks at $\Delta \approx 0$, and the dark exciton density exhibits a sharp nonlinear rise, which is manifested as a sudden blueshift of the PL.

Hysteresis behavior, a hallmark of bistability, is clearly seen in Fig. 4a, which displays the IX energy for increasing (blue) and decreasing (orange) pump power [29,32-34]. Remarkably, the blueshift onset when ramping the power up, $P_{up}$, is an order of magnitude larger than when ramping it down, $P_{down}$. In Fig. 4b, we fix the pump power just above $P_{up}$ at $B = 0$ and ramp the magnetic field up to 1 T, and then back to zero. The hysteresis behavior as a function of $B$ confirms the buildup of nuclear polarization: The magnetic field increases $\Delta_0$, requiring larger $P_N$ to screen it, and the bistability loop is affected accordingly.

Let us turn to showing that $\Delta \approx 0$ at threshold. The electron Zeeman energy $\approx$ 10 $\mu$eV/T, which defines $\Delta_0$, is much smaller than the PL linewidth in the relevant magnetic field range. Since this value is expected to dramatically reduce by the Overhauser field, it cannot be resolved by straightforward spectroscopic measurements. Hence, to determine the value of $\Delta$ we conducted optically detected RF measurement, where an oscillating magnetic field co-polarized with $B$, is applied. The hyperfine interaction couples the electron and nuclear spins such that the transition energy between the coupled electron-nuclear spin states is $\approx \Delta + h\nu_N^i$, where $h\nu_N^i$ is the $i$-isotope Zeeman energy. When the RF frequency, $\nu_{RF}$, is resonant with this transition, there is an efficient transfer of RF power to the electron system, and the optical threshold power is reduced [see section 4 of the SM]. Hence, by measuring the resonance frequency at which $P_{up}$ or $P_{down}$ are minimal at a given $B$, and using the known value for $\nu_N^i$, we can extract $\Delta$.

Figure 5a shows $P_{up}$ as a function of the RF frequency at $B = 1$T. Three narrow resonances, in which threshold drops to a lower value, are clearly resolved. The frequencies of these resonances are close to the expected nuclear Zeeman frequencies, $\nu_N^i = \gamma_N^i B$, of the three isotopes in GaAs, $^{75}$As, $^{69}$Ga, and $^{71}$Ga [24,26]. This is demonstrated in Fig. 5b, which shows that the evolution of the lowest resonance with magnetic field closely follows the nuclear Zeeman frequency curve for $^{75}$As, denoted by the dashed line. Furthermore, the relative strength of the resonances is in good agreement with the abundance of these isotopes, $\sim 100{:}\,60{:}\,40$ [24,26]. We can therefore associate these three resonances with transitions involving the corresponding nuclear isotopes. The blue and red circles in Fig. 5c show the frequency difference, $\delta\nu = \nu_{RF}^i - \nu_N^i$, for the $^{75}$As and $^{69}$Ga resonances, respectively. We find the same values of $\delta\nu$ for both isotopes, confirming that the frequency difference is not isotope specific, but is rather due to the residual Zeeman energy of the electron, $h\delta\nu = \Delta$. Comparing the value of $\delta\nu$ at 1T, 0.8 MHz, to the bare electron Zeeman frequency for a 12 nm QW [35], $\Delta_0 \sim 3$ GHz, we obtain $\Delta/\Delta_0 \sim 2.5 \times 10^{-4}$, demonstrating the extent to which the Zeeman gap is screened by the Overhauser field. We show in the Supplementary material section that the residual gap, $\delta\nu$, is due to fluctuations in the number of polarized nuclei within an exciton volume. At small magnetic fields, only a small fraction of the nuclei in the sample are polarized at steady state, while the rest are randomly oriented with a zero mean. An exciton, which samples a finite volume of nuclei, would therefore be subjected to a random nuclear field due to the finite size sampling [see section 5 of the SM].

This strong nonlinear coupling between the condensate and nuclear spins stabilizes the interference pattern and allows its observation. The density modulation near the boundary is imprinted in the slowly varying nuclear polarization and thereby yields a modulation of the PL intensity. Indeed, we find that the interference pattern builds over seconds upon turning on the pump beam and is observed only beyond the second threshold, where the nuclei are at the highly polarized state, $\Delta \approx 0$.

**Conclusion**

In the concluding part we wish to highlight several unique properties of the dark condensate that should stimulate further investigations. Similarly to cavity exciton-polaritons, the system constitutes a non-equilibrium, driven-dissipative coherent state. However, owing to the dark exciton larger effective mass and extended lifetime, it exhibits a more matter-like character that is close to equilibrium BEC, thereby accessing a qualitatively different region of parameter space. The intrinsic large electrical dipole moment of the excitons enables external control through electrostatic gating, evidenced, for example, by the boundary reflection phenomena, and may also allow probing via coupling to electrical currents [8]. Finally, the emergence of nuclear polarization across the entire mesa, mediated collectively by a mobile quantum fluid and exhibiting locking at $\Delta \approx 0$, suggests intriguing prospects for quantum control and manipulation.

**Data and materials availability**: All data needed to evaluate the conclusions in the paper are present in the paper and/or the Supplementary Materials.

**Acknowledgments:** We wish to thank Michael Stern, Subhradeep Misra, and Alexander Poddubny for fruitful discussions.

**Funding:** This work is supported by the Israeli Science Foundation, grant 714922 (to I.B.J.).

**References:**

1. Yu. E. Lozovik and O. L. Berman, Phase transitions in a system of spatially separated electrons and holes, JETP 84 1027 (1997).
2. L. V. Butov, A. C. Gossard, D. S. Chemla, Macroscopically ordered state in an exciton system. Nature **418**, 751-754 (2002).

3. D. Snoke, S. Denev, Y. Liu, L. Pfeiffer, K. West, Long-range transport in excitonic dark states in coupled quantum wells. *Nature* **418**, 754-757 (2002).
4. L. V. Butov, Condensation and pattern formation in cold exciton gases in coupled quantum wells, *J. Phys.: Condens. Matter* **16** R1577 (2004).
5. B. Laikhtman, R. Rapaport, Exciton correlations in coupled quantum wells and their luminescence blue shift. *Phys. Rev. B* **80**, 195313 (2009).
6. A. A. High, J. R. Leonard, A. T. Hammack, M. M. Fogler, L. V. Butov, A. V. Kavokin, K. L. Campman, A. C. Gossard, Spontaneous coherence in a cold exciton gas. *Nature* **483**, 584-588 (2012).
7. D. Snoke, Dipole Excitons in Coupled Quantum Wells: Towards an Equilibrium Exciton Condensate, in: R. G. Hulet, A. G. Truscott, and K. A. Gibble (eds.), *Quantum Gases: Finite Temperature and Non-Equilibrium Dynamics*, World Scientific, Chapter 28, 751 (2013).
8. M. Stern, V. Umansky, I. B. Joseph, Exciton liquid in coupled quantum wells. *Science* **343**, 55-57 (2014).
9. S. Misra, M. Stern, A. Joshua, V. Umansky, I. B. Joseph, Experimental Study of the Exciton Gas-Liquid Transition in Coupled Quantum Wells, *Phys. Rev. Lett.* **120**, 047402 (2018).
10. Z. Wang, D. A. Rhodes, K. Watanabe, T. Taniguchi, J. C. Hone, J. Shan, K. F. Mak, Evidence of high-temperature exciton condensation in two-dimensional atomic double layers. *Nature* **574**, 76-80 (2019).
11. M. Combescot, O. Betbeder-Matibet, R. Combescot, Bose-Einstein condensation in semiconductors: The key role of dark excitons. *Phys. Rev. Lett.* **99**, 176403 (2007).
12. R. Combescot, M. Combescot, "Gray" BCS condensate of excitons and internal Josephson effect. *Phys. Rev. Lett.* **109**, 026401 (2012).
13. Y. Shilo, K. Cohen, B. Laikhtman, K. West, L. Pfeiffer, R. Rapaport, Particle correlations and evidence for dark state condensation in a cold dipolar exciton fluid. *Nat. Comm.* **4**, 2335 (2013).
14. M. Alloing, M. Beian, M. Lewenstein, D. Fuster, Y. González, L. González, R. Combescot, M. Combescot. F. Dubin Evidence for a Bose-Einstein condensate of excitons. *Europhys. Lett*. **107**, 10012 (2014).

15. M. Beian, M. Alloing, E. Cambril, C.armen G. Carbonel, J. Osmond, A. Lemaître, F. Dubin, Long-lived spin coherence of indirect excitons in GaAs coupled quantum wells. *Europhys. Lett.* **110**, 27001 (2015).
16. K. Cohen, Y. Shilo, K. West, L. Pfeiffer, R. Rapaport, Dark high density dipolar liquid of excitons. *Nano Lett*. **16**, 3726-3731 (2016).
17. M. Combescot, R. Combescot, F. Dubin, Bose-Einstein condensation and indirect excitons: A review. *Rep. Prog. Phys*. **80**, 066501 (2017).
18. Y. Mazuz-Harpaz, K. Cohen, M. Leveson, K. West, L. Pfeiffer, M. Khodas, R. Rapaport, Dynamical formation of a strongly correlated dark condensate of dipolar excitons. *Proc. Natl. Acad. Sci. U.S.A*. **116**, 18328-18333 (2019).
19. S. Misra, M. Stern, V. Umansky, I. B. Joseph, The role of spin-flip collisions in a dark-exciton condensates. *Proc. Natl. Acad. Sci*. **119**, e2203531119 (2022).
20. A. Jash, M. Stern, S. Misra, V. Umansky, I. B. Joseph, Giant hyperfine interaction between a dark exciton condensate and nuclei. *Sci. Adv.* **10**, eado8763 (2024).
21. N. W. Sinclair, J. K. Wuenschell, Z. Vörös, B. Nelsen, and D. W. Snoke M. H. Szymanska, A. ChinJ. Keeling, L. N. Pfeiffer, K. W. West, Strain-induced darkening of trapped excitons in coupled quantum wells at low temperature. *Phys. Rev. B* **83**, 245304 (2011).
22. I. Carusotto, C. Ciuti, Quantum fluids of light. *Rev. Mod. Phys*. **85**, 299–366 (2013).
23. V. N. Mantsevich, M. M. Glazov, Viscous hydrodynamics of excitons in van derWaals heterostructures. *Phys. Rev. B* **110**, 165305 (2024).
24. D. Paget, G. Lampel, B. Sapoval, V. I. Safarov, Low field electron-nuclear spin coupling in gallium arsenide under optical pumping conditions. *Phys. Rev. B* **15**, 5780-5796 (1977).
25. V. K. Kalevich, K. V. Kavokin, I. A. Merkulov, *"Dynamic nuclear polarization and nuclear fields"* in Spin Physics in Semiconductors, Mikhail I. Dyakonov editor (Springer Berlin, Heidelberg, 2008).
26. M. Dobers, K. v. Klitzing, J. Schneider, G. Weimann, K. Ploog, Electrical detection of nuclear magnetic resonance in GaAs−$Al_xGa_{1-x}As$ heterostructures. *Phys. Rev. Lett.* **61**, 1650–1653 (1988).
27. D. Gammon, Al. L. Efros, T. A. Kennedy, M. Rosen, D. S. Katzer, D. Park, S. W. Brown, V. L. Korenev, I. A. Merkulov, Electron and nuclear spin interactions in the optical spectra of single GaAs quantum dots. *Phys. Rev. Lett.* **86**, 5176-5179 (2001).

28. I. A. Merkulov, Al. L. Efros, M. Rosen, Electron spin relaxation by nuclei in semiconductor quantum dots. *Phys. Rev. B* **65**, 205309 (2002).
29. P. Maletinsky, C. W. Lai, A. Badolato, A. Imamoglu, Nonlinear dynamics of quantum dot nuclear spin. *Phys. Rev. B* **75**, 035409 (2007).
30. A. I. Tartakovskii, T. Wright, A. Russell, V. I. Fal'ko, A. B. Van'kov, J. Skiba-Szymanska, I. Drouzas, R. S. Kolodka, M. S. Skolnick, P. W. Fry, A. Tahraoui, H.-Y. Liu, M. Hopkinson, Nuclear spin switch in semiconductor quantum dots. *Phys. Rev. Lett.* **98**, 026806 (2007).
31. $\Delta_0$ is the sum of Zeeman and exchange energy.
32. E. Aubay, D. Gourier, Magnetic bistability and Overhauser shift of conduction electrons in gallium oxide. *Phys. Rev. B.* **47**, 15023 (1993).
33. P.-F. Braun, B. Urbaszek, T. Amand, X. Marie, O. Krebs, B. Eble, A. Lemaitre, P. Voisin, Bistability of the nuclear polarization created through optical pumping in $In_{1-x}Ga_xAs$ quantum dots. *Phys. Rev. B.* **74**, 2453063 (2006).
34. T. Belhadj, T. Kuroda, C.-M. Simon, T. Amand, T. Mano, K. Sakoda, N. Koguchi, X. Marie, B. Urbaszek, Optically monitored nuclear spin dynamics in individual GaAs quantum dots grown by droplet epitaxy. *Phys. Rev. B.* **78**, 205325 (2008).
35. A. V. Shchepetilnikov, Y. A. Nefyodov, I. V. Kukushkin, W. Dietsche, Electron g-factor in GaAs/AlGaAs quantum wells of different width and barrier Al concentrations. *J. Phys. Conf. Ser*. **456** 012035 (2013).

**Figure:**

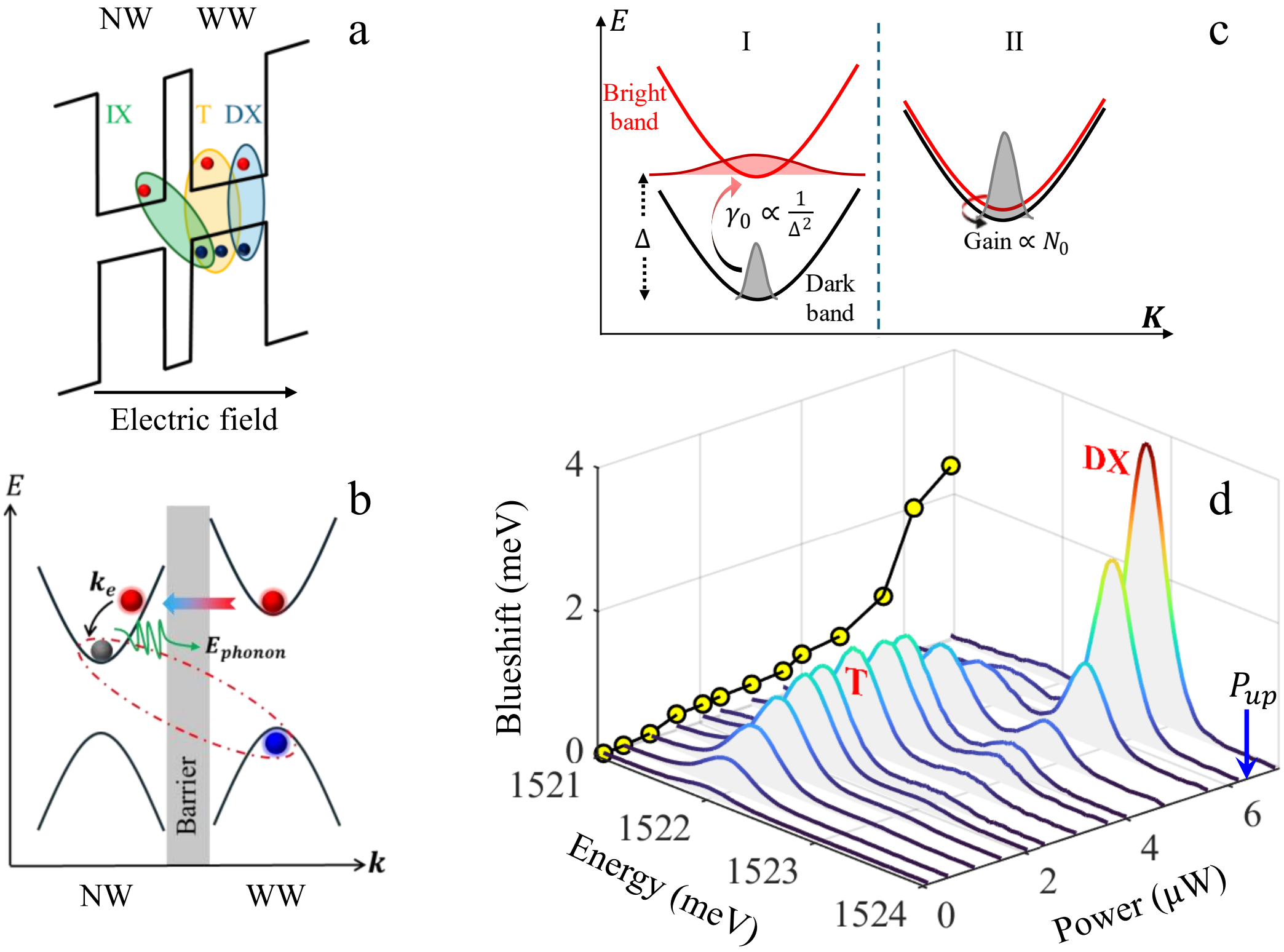

Fig. 1. a. The coupled quantum wells (CQWs) under electric field and the three bound states that are formed under optical excitation in the wide well (WW): Direct exciton (DX), trion (T), and indirect exciton (IX). b. Schematic of IX formation: Hot electrons generated in the WW tunnel into the narrow well (NW), and relax via phonon emission. c. Panel I: At low power the dark band is only weakly coupled to the bright band, and condensation occurs at the dark band minimum. Panel II: At high power the Overhauser shift closes the dark-bright gap, and bright → dark spin-flip transitions are strongly enhanced. d. The evolution of the photoluminescence (PL) with excitation power reveals the disappearance of the trion line above a threshold power, $P_{up}$, together with an enhanced blueshift of the IX line (yellow circles). Here, $T$ = 1.5 K.

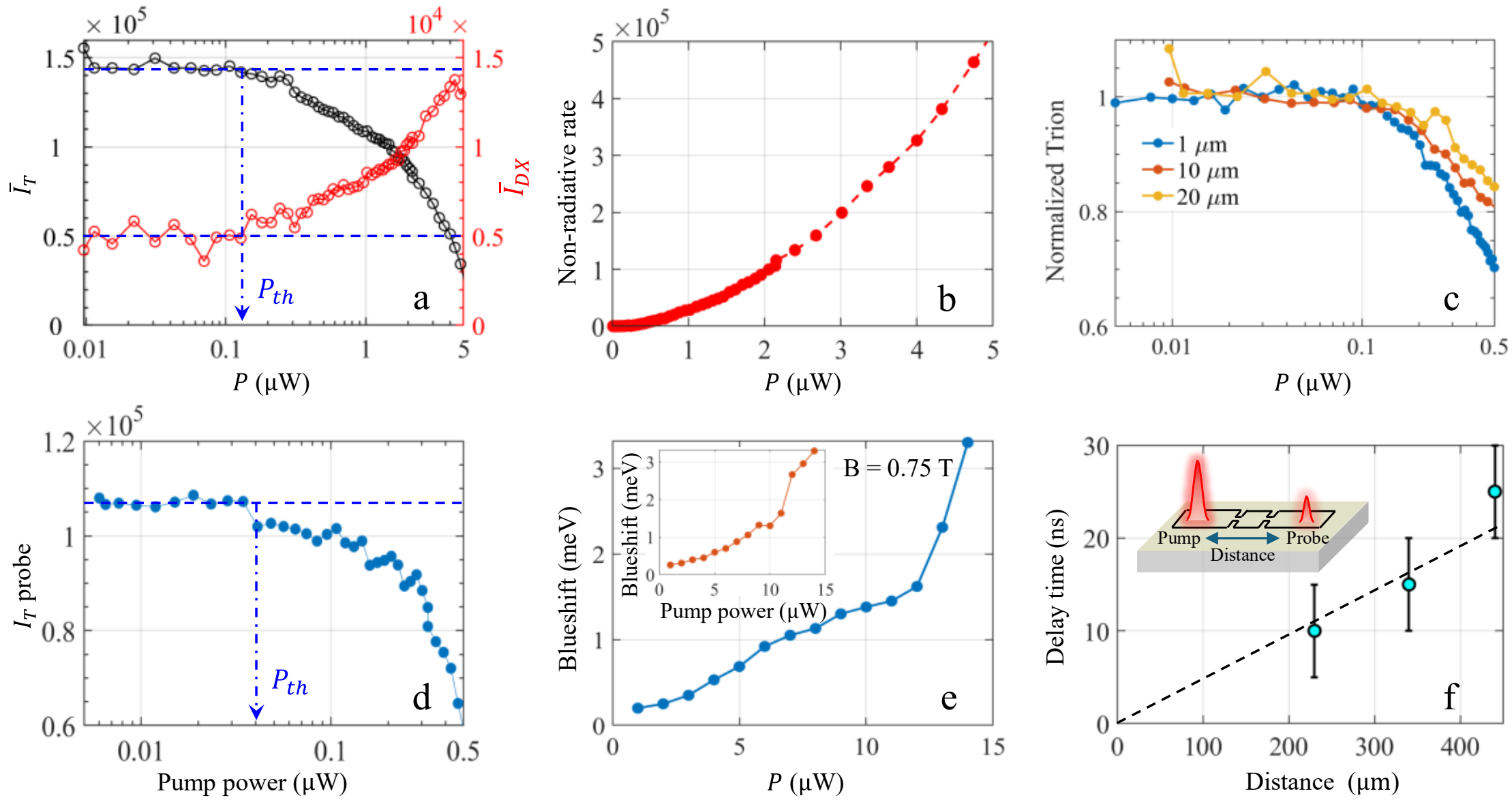

Fig. 2. a. The normalized trion and DX intensity, $\bar{I}_T$ and $\bar{I}_{DX}$, respectively as a function of power in units of $\text{counts}/(\text{sec} \cdot \mu\text{W})$. Blue dashed arrow indicates the threshold power for condensation, $P_{th}$. b. Non-radiative recombination rate as a function of power in units of $\text{counts}/\text{sec}$. c. The normalized trion intensity as a function of power for different beam spot size, showing that $P_{th}$ is independent of excitation intensity. Trion intensities are normalized with the low power value. d. The trion intensity at the probe location as function of pump power ($T = 0.6$ K). The threshold power is lower than in Fig. a due to the finite probe power, 50 nW. e. IX energy shift (blueshift) versus excitation power at $B = 0.75$T. The magnetic field increases the dark-bright splitting, enabling measurement of the linear IX blueshift over a wide power range. Inset: a similar blueshift is observed at the probe position when the pump is placed far from the probe, indicating that the dark condensate extends across the entire mesa. f. The delay time between the pump and various probe positions ($T = 0.6$ K). The slope of black dashed line corresponds to the expected sound velocity of a dipolar exciton liquid, $\sim 2 \times 10^4$ m/s, for a density of $\sim 10^{10}$ cm$^{-2}$. The inset shows a schematic of the pump-probe measurement geometry.

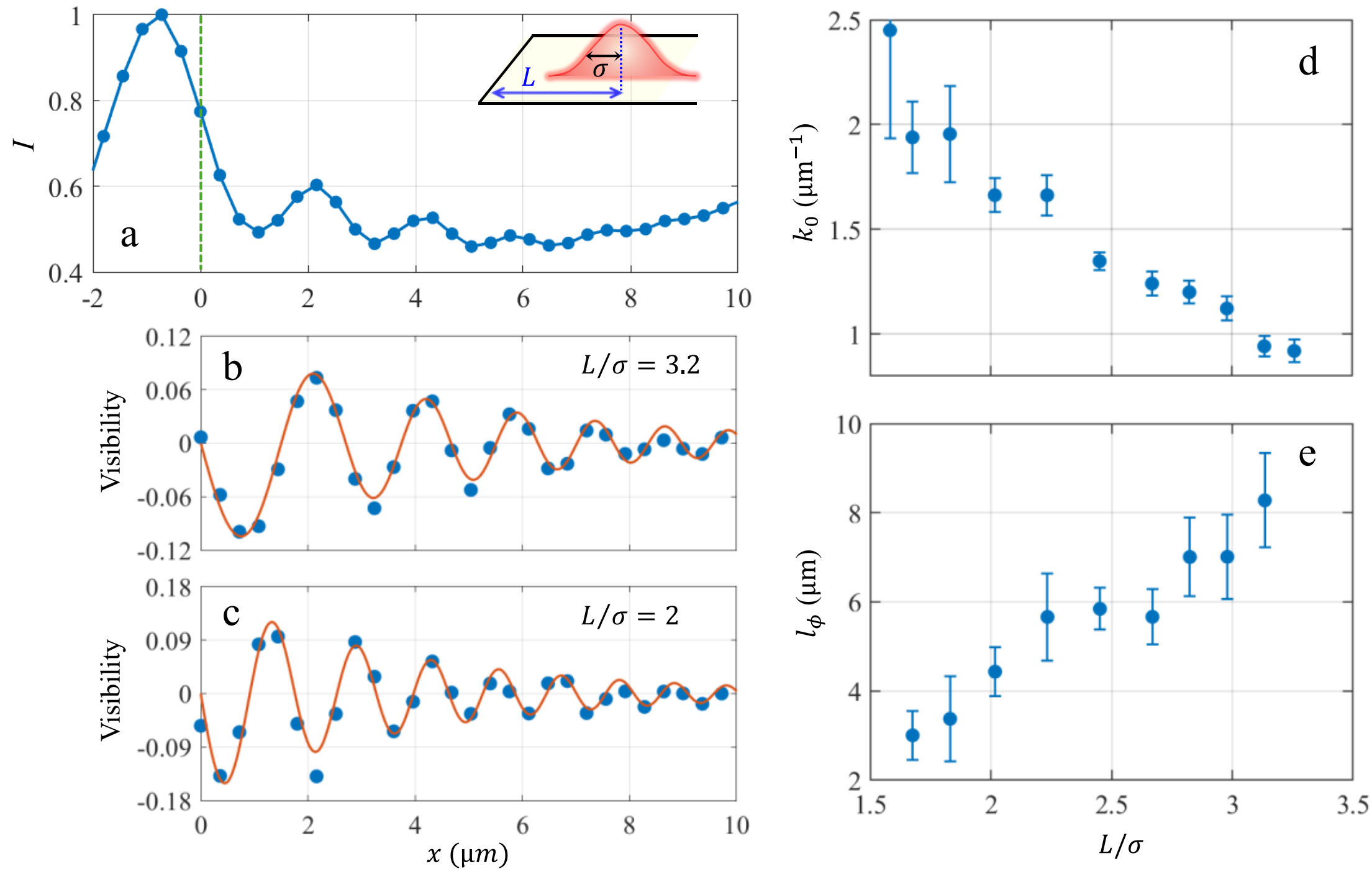

Fig. 3. a. The cross-section of PL intensity, $I(x)$, when illuminated close to the sample edge. The green dashed line indicates the position of the top-gate edge. The strong PL to the left of the green line is due to scattered PL from the edge of the CQW. Inset: schematic of the measurement geometry, where $L$ is the distance between the excitation spot center and the gate edge, and $\sigma$ is the Gaussian beam radius. b. The normalized background subtracted visibility curve, $(I-\bar{I})/\bar{I}$, of the interference fringes extracted from Fig. 3a, for $L/\sigma = 3.2$ (here $\bar{I}$ is the local mean background PL). The red solid line shows a fit to $e^{-x/l}\sin(2kx)$, with $k = k_0 + \alpha x$, demonstrating the chirping of the interference pattern and decaying envelope. c. Visibility curve for a different excitation position with $L/\sigma = 2$, clearly showing the change in periodicity and decay length. d. Extracted values of $k_0$ as a function of $L/\sigma$. e. The coherence length $l_\phi$, extracted from the envelope decay, versus $L/\sigma$. In a reflection geometry, the measured decay length, $l$, corresponds to $l_\phi/2$.

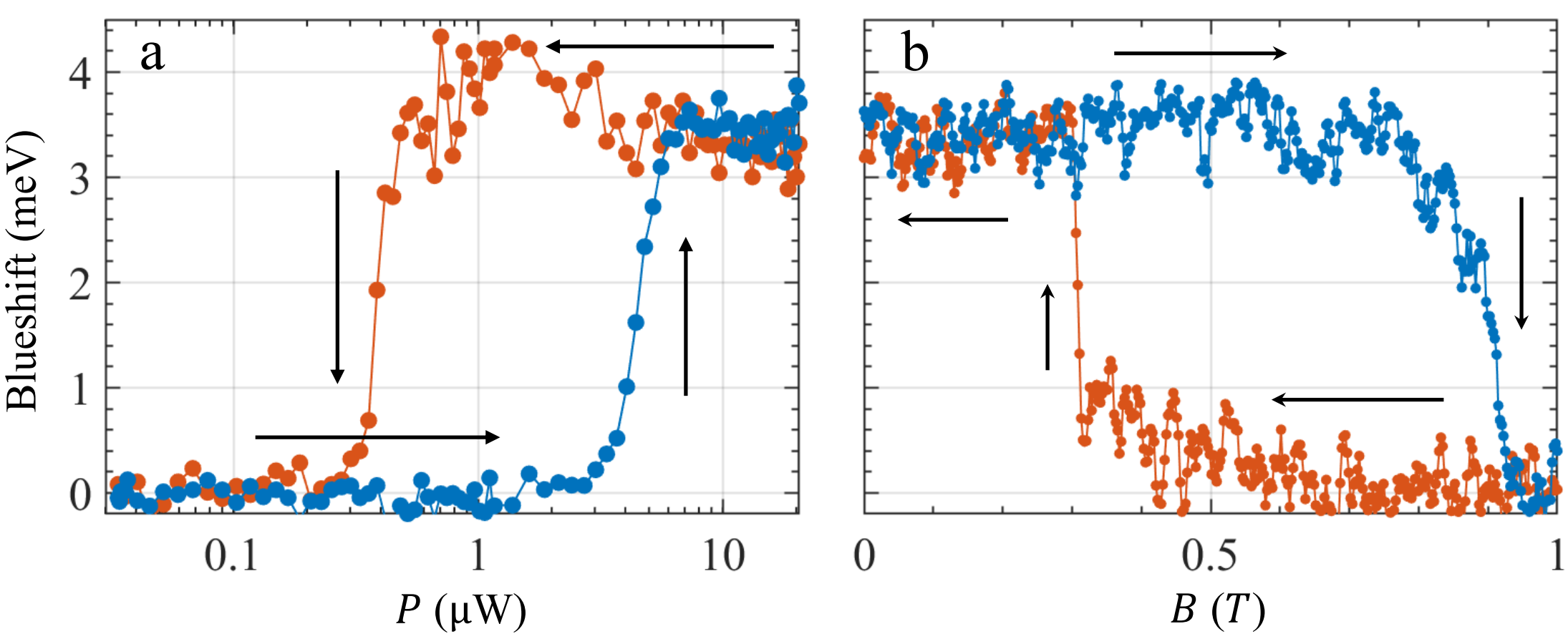

Fig. 4. a. The blueshift as a function of ramping the power up and down at $B = 0$ T. b. The blueshift as a function of magnetic field at constant laser power of 5.3 μW. Here, $T = 1.5$ K.

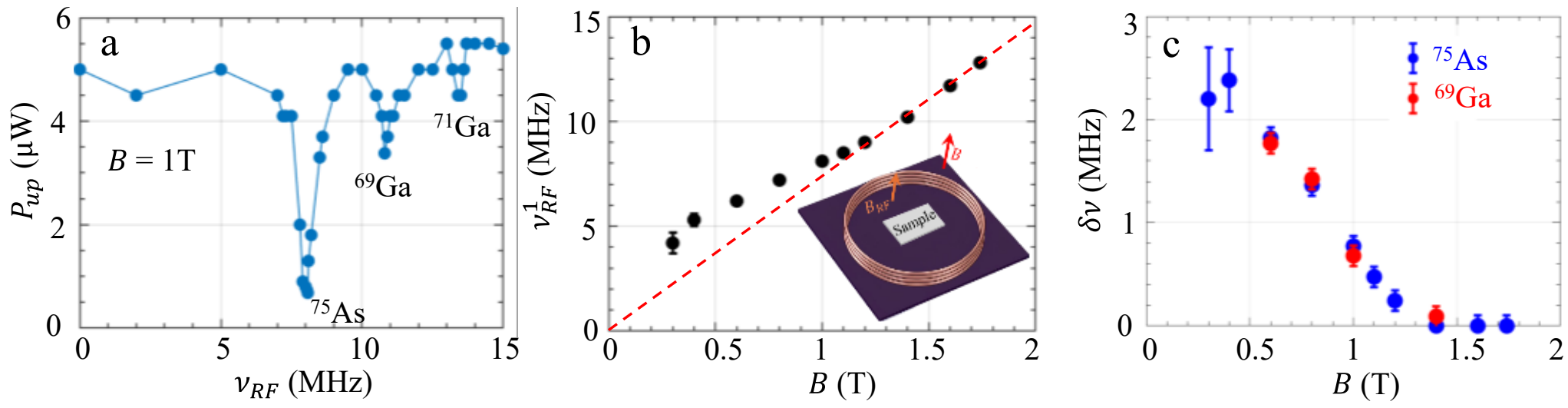

Fig. 5. a. The threshold power when the power is ramped up at 1 T as a function of $\nu_{RF}$ . The three resonances at 8.0, 10.9, and 13.5 MHz are clearly observed. b. The resonance frequency of $^{75}$As (scattered black points) as a function of magnetic field. The red dashed line corresponds to $\gamma_N^1 B$ for $^{75}$As, where $\gamma_N^1 = 7.295$ MHz/T. Inset shows the RF measurement setup, showing the external DC magnetic field, $B$, and the coil that generates the oscillating magnetic field, $B_{RF}$. c. The scattered data show $\delta\nu$ as a function of $B$ for two isotopes, $^{75}$As and $^{69}$Ga. Here, $T = 1.5$ K.

# **Coherence, long-range transport and nuclear polarization in a driven-dissipative dark exciton condensate**

Amit Jash, Maheswar Swar, Uri Shimon, Vladimir Umansky, and Israel Bar-Joseph

*Department of Condensed Matter physics, Weizmann Institute of Science, Rehovot 7610001, Israel*

## Contents

# 1. Sample details and methods

A schematic of the sample and its constituting layers are shown in Fig. S1. The coupled quantum wells (CQWs) system is the same as in [1] and consists of two GaAs quantum wells having widths of 12 and 18 nm that are separated by a 3 nm wide $Al_{0.28}Ga_{0.72}As$ barrier (see Fig. S1(a)). It is embedded in a 2 μm wide $n-i-n$ structure so that a voltage can be applied in the perpendicular direction. The top and bottom $n+$ layers are silicon doped ($n_{Si}$ ~ $10^{18}$ $cm^{-3}$) $Al_{0.12}Ga_{0.88}As$. The intrinsic region consists of two superlattice (SL) layers, each 1 μm thick: below the CQW, we have 33 periods of [27 nm $Al_{0.37}Ga_{0.63}As$] – [2 nm AlAs] – [1 nm GaAs], and above the CQW, 20 periods of [50 nm $Al_{0.33}Ga_{0.67}As$] – [1 nm GaAs] (see Fig. S1(b)). This SL structure ensures that the crystal quality is maintained, as indeed reflected in the PL linewidth of the two wells, which is sub-meV. The direct-exciton linewidth from our sample is approximately $0.60 \pm 0.08$ meV. The band gap of the contact layers, 1.650 eV, and the SL layers, ~1.9 eV, is selected to be well above the illumination energy of the laser so that these layers do not contribute to carrier photo-generation. The sample structure was grown in our MBE system and patterned into mesas by optical lithography, wet etching, and metal deposition by thermal evaporation. The experiments reported in this paper were conducted on three different mesas with dimensions (500 ×100 $\mu m^2$ mesa) and were found to be highly reproducible.

Upon application of a high enough negative voltage (below $-1$V) to the top gate, the electron level in the wide well (WW) becomes higher than the corresponding level in the narrow well (NW). This enables electrons to tunnel from the WW to the NW (refer to Fig. 1a in the main manuscript) and form indirect excitons.

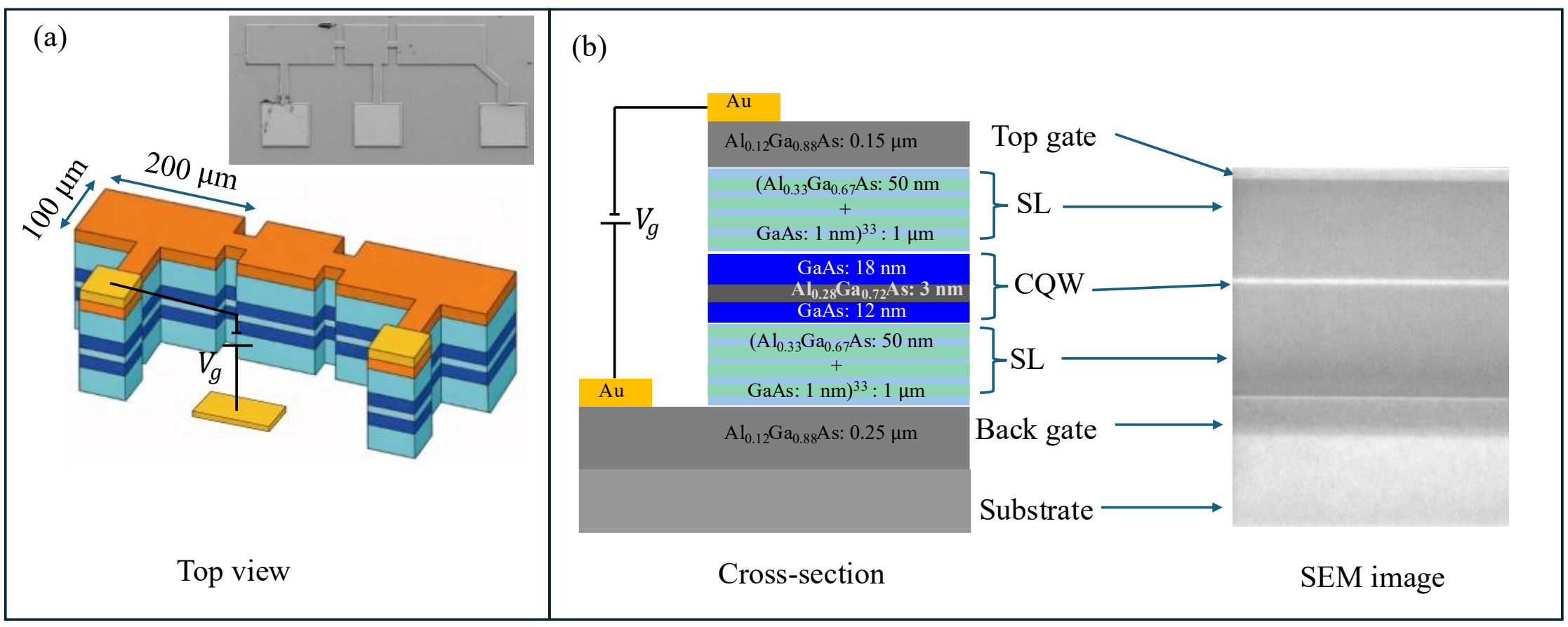

Fig. S1. (a) Schematic of the sample structure; the inset shows a SEM image of the mesa and the top gate. (b) Layer-growth details and an SEM cross-sectional view of the device.

*Sample edges:* Fig. S2 shows an AFM line profile (a) and an SEM image (b) of the mesa boundary. As can be seen, the mesa boundary is highly anisotropic due to different etching rates along orthogonal crystalline orientations: a high etch rate along the y-axis (vertical edge in Fig. S2(b)) and a slower rate along the x-axis (horizontal edge). The y-boundary develops a faceted, pyramid-like profile; therefore, excitons approaching this side experience a smooth, repulsive high potential and are reflected before reaching the physical CQWs boundary. An interference pattern is observed near this mesa boundary when the excitation laser beam is focused close to this. In contrast, the horizontal boundary remains sharp, such that exciton reflection occurs predominantly at the physical CQW edge. In this case, the CQW edge is defined by the lossy etched surface, which quenches the interference. Excitons approaching this boundary experience non-radiative interface loss and dephasing, and are absorbed there. Therefore, no interference pattern is observed at this boundary.

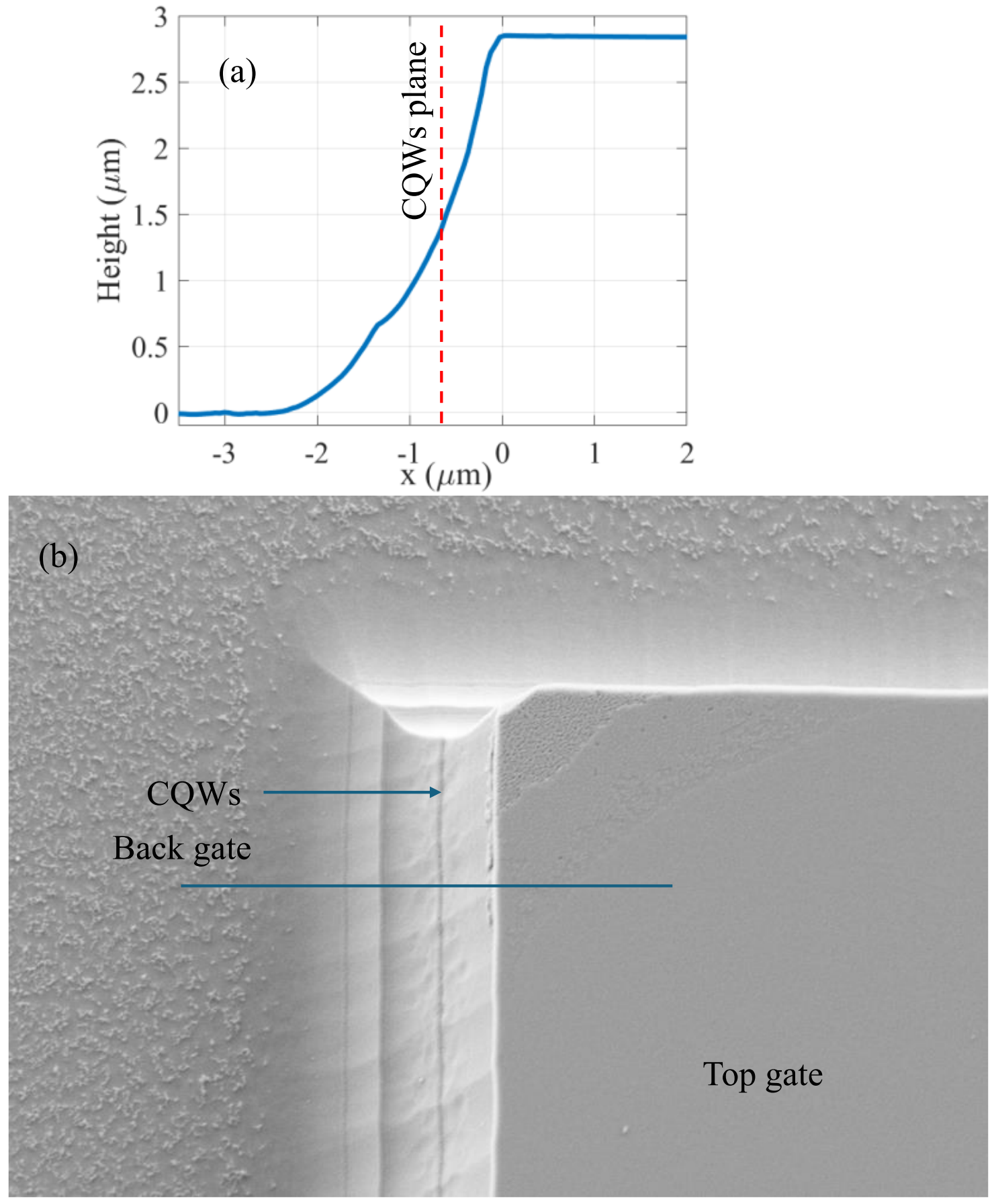

Fig. S2: An AFM line profile (a) and SEM image (b) of the mesa boundary of the sample.

**Methods:** The data presented in this paper were collected using three different cryostats: (I) a dry refrigerator (attoDRY2100) integrated with a confocal microscopy setup, operating at a base temperature of 1.7 K and magnetic fields up to 9 T, (II) a dilution refrigerator with optical access, covering a temperature range from 100 mK to 5 K, and (III) a split-coil, magneto-optical pumped-helium cryostat, operating between 1.5 K and 10 K with magnetic fields up to 7 T.

A schematic of the optical setup is shown in Fig. S4. The sample is illuminated by a Ti:sapphire laser through a high-NA objective placed inside the cryostat, with the sample positioned at the focal plane. The laser energy is 1.528 meV, and the Gaussian spot size on the sample varies between 1 and 40 μm for various measurements. The laser energy is adjusted below the NW energy, such that the carriers are only created in the WW. The resulting photoluminescence (PL) is collected by the same objective. The collected signal is then directed to a Spex (500 M) spectrometer and an Andor iXon EMCCD camera for spectral and imaging measurements, respectively.

In the PL collection path, unwanted reflected laser light from different parts of the optical setup is suppressed by more than five orders of magnitude using a polarizer oriented perpendicular to the incident laser polarization, together with a high-quality optical filter. A 50:50 beam splitter is used to separate the excitation and collection paths. A pair of lenses (telescope configuration) is used to focus the PL, and a pinhole mounted on a precision three-axis translation stage is employed to selectively collect PL from different regions of the illumination area.

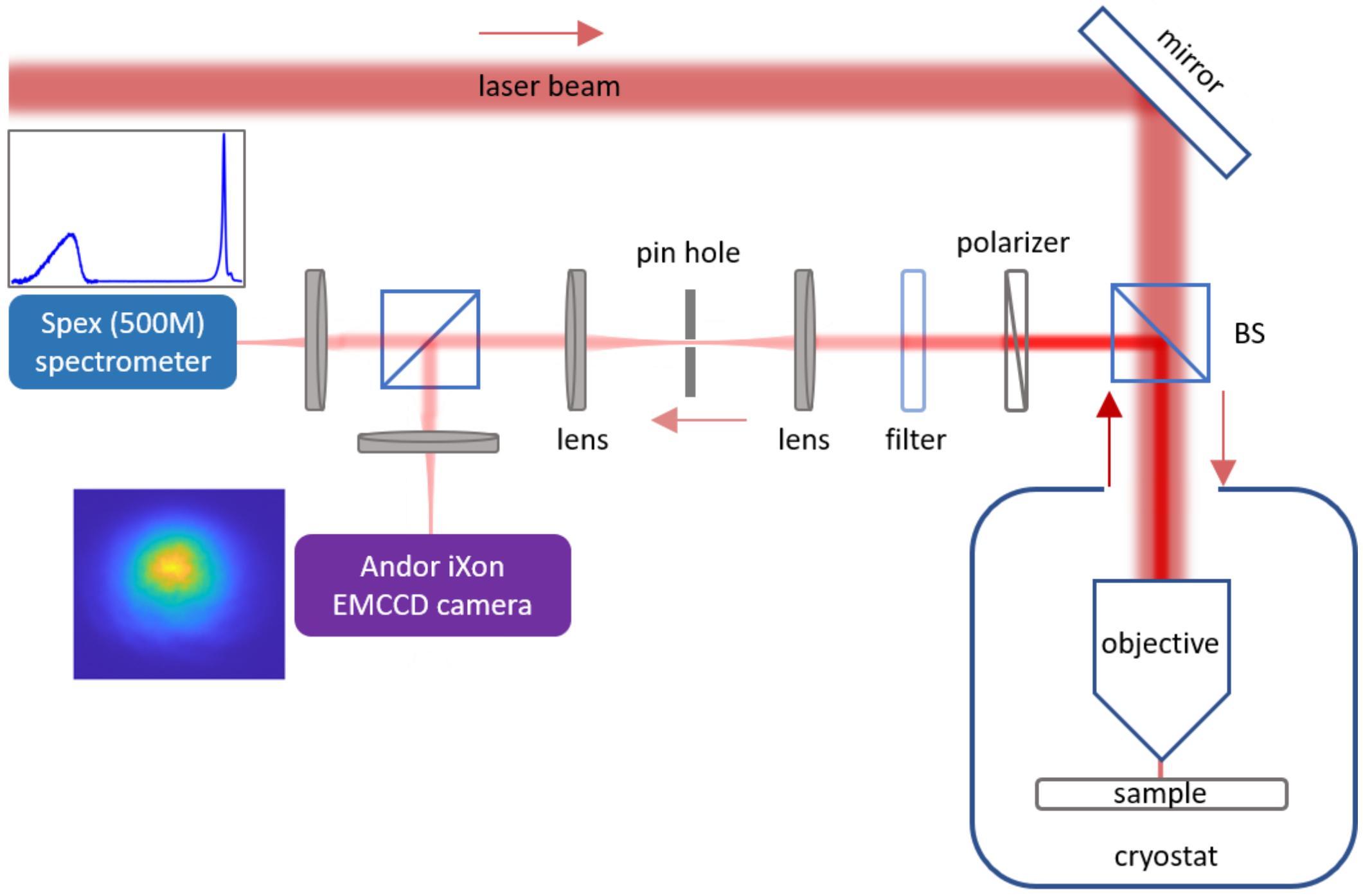

Fig. S3: Schematic of the experimental setup.

Pump-probe measurements are performed in cryostats (II) and (III). The probe beam is generated by a diode laser with a similar spot size and an energy of 1.5794 eV. The pump spot is fixed on the sample, while the probe position is varied using a precision translation stage. PL from selected regions of the mesa is collected using a pinhole placed at the focal plane of the telescope in the detection path. A low-noise Keithley 2400 source meter is used to apply the gate voltage.

For the RF measurements, we have built a coil above the 24-pin chip holder. An RF signal is applied through the coil using a DS345 function generator, generating oscillatory RF magnetic field around the sample along z-direction. To determine $P_{th}$ at a specific frequency, we adjusted the RF radiation to a range between 0 and 30 MHz and measured the PL spectrum while slowly increasing the power, with a 5-second pause between each measurement.

## 2. Condensation at the dark band

In general, a driven-dissipative condensate can be reduced into a two-manifold model, a pumped reservoir that feeds a single low-$K$ mode ("ground state"). High occupation of that mode yields stimulation into this ground state above a certain threshold. Let us define $n_R$ - the reservoir density, $\gamma_R$ - the reservoir loss rate, $R$ - the scattering rate into the low-$K$ mode, $\psi(r) = \psi_0 \exp(iKr)$ - the ground state mode, $\gamma_0$ - the total loss from that mode, $g$ - the exciton-exciton interaction, and $P$ - the pumping rate into the reservoir. Then the rate equation for the reservoir is,

$$\frac{dn_R}{dt} = P - (\gamma_R + R|\psi|^2)n_R. \tag{1}$$

and the generalized Gross-Pitaevskii equation (GPE) [2] is,

$$\frac{d}{dt}|\psi|^2 = \left[-\frac{\hbar^2}{2m}\nabla^2 + g|\psi|^2 + i(Rn_R - \gamma_0)\right]|\psi|^2. \tag{2}$$

In the first order, the ground state occupation rate can be considered as $Rn_R$ [2]. Rewriting using the occupation of the ground mode $N_0 = |\psi_0|^2$, we obtain two rate equations,

$$\begin{aligned} \frac{dN_0}{dt} &= Rn_R(1 + N_0) - \gamma_0 N_0, \\ \frac{dn_R}{dt} &= P - [R(1 + N_0) + \gamma_R]\, n_R. \end{aligned}$$

Threshold is when $Rn_R > \gamma_0$. With $n_R \approx \frac{P}{\gamma_R}$ below threshold, $P_{\text{th}} = \frac{\gamma_0\gamma_R}{R}$. Above threshold, $N_0$ grows and clamps the reservoir (“gain clamping”), just like a laser, but unlike laser the gain comes from Bose stimulation $(1 + N_0)$, not inversion.

### a. Implementation to CQW

Let us consider the following route for condensation in our CQW, consisting of a wide (WW) and narrow well (NW). We apply electric field, such that the electron level in the NW is lower than that in the WW, and electrons may tunnel to the NW. Since the excess energy of the tunneling electrons is large, a few to tens of meV, the electrons are initially hot. The capture of a hot NW electron, $e_{NW}$, with in-plane momentum $\boldsymbol{k}_e$ and a WW cold hole, $h_{WW}$, into an IX 1s state with center of mass (CM) momentum $\boldsymbol{K}$ is achieved by emitting one (or more) acoustic phonon with momentum $\boldsymbol{q}$. The phonon magnitude, $q = \sqrt{q_\parallel^2 + q_z^2}$, is fixed by the hot electron energy, $\Delta E = \hbar v_s q$, where $v_s$ is the speed of sound in GaAs. Its direction, however, is set by in-plane momentum conservation $\boldsymbol{k}_e - \boldsymbol{K} = \boldsymbol{q}_\parallel$, where $|\boldsymbol{q}_\parallel| \le q$. The rate of this process can be written as,

$$R(q) \propto \left|M_{e-\mathrm{ph}}(q)\right|^2 \left|F_{1s}(q_\parallel)\right|^2 \left|F_z(q_\parallel)\right|^2 \delta(\hbar v_s q - \Delta E). \qquad (3)$$

Here, $M_{e-\mathrm{ph}}(q)$ is the electron-phonon matrix element (deformation potential, piezo-electric), $F_{1s,z}$ are the in plane (into the exciton 1s orbital) and out of plane form factors.

**b. Calculation of the $\boldsymbol{q}$ dependence of the form factors $\boldsymbol{F_{1s}, F_z}$**

Let the e-h final state be,

$$\Phi_{1s,K}(\boldsymbol{\mathcal{R}}, \rho) = e^{i\mathbf{K}\cdot\boldsymbol{\mathcal{R}}}\, \phi_{1s}(\rho)\, \chi_e(z)\chi_h(z).$$

where $\phi_{1s}(\rho)$ is the 2D 1s exciton relative wavefunction, $\phi_{1s}(\boldsymbol{\rho}) = \frac{1}{\sqrt{2\pi}a_B}\exp\left(-\frac{\rho}{a_B}\right)$, $\boldsymbol{\mathcal{R}}$ is the CM coordinate, $a_B$ is the Bohr radius and $\chi_{e,h}(z)$ are the $e_{NW}, h_{WW}$ wavefunctions. Projection of the initial free $e_{NW}, h_{WW}$ states on the final state $\Phi_{1s,K}$ yields two form factors:

- In-plane (relative-motion) form factor $\mathrm{F_{1s}}$

$$F_{1s}(q) = \int d^2\rho\, e^{-i\mathbf{q}\cdot\boldsymbol{\rho}} \phi_{1s}(\rho) = \sqrt{2\pi}\, a_B\, \frac{1}{(1 + a_B^2 q_\parallel{}^2)^{\frac{3}{2}}}.$$

- Vertical (interlayer) form factor $\mathrm{F_z}$

$$F_z(q) = \int dz \int dz'\, |\chi_e(z)|^2 |\chi_h(z')|^2\, e^{-q|z-z'|}.$$

In the Delta-layer approximation and well centers separated by $d$:

$$F_z(q) \simeq e^{-qd}.$$

For the electron-phonon matrix elements we also get a $z$-form factor, $I_w$,

$$I_w(q_z) = \left| \int dz\, |\chi_w(z)|^2 e^{iq_z z} \right|^2.$$

So, the transition rate by phonon emission can be written as

$$R(q) \propto \exp\,[-2q_{\parallel}d]\ [1 + (\frac{a_B q_{\parallel}}{2})^2]^{-3}\ I_w(q_z)\,.$$

This is a decreasing function of $q$: the two first factors drop with $q_{\parallel}$ and the function $I_w(q_z)$ penalizes large vertical phonon momentum $q_z$.

### c. Scattering of hot electrons with a distribution $\boldsymbol{f_e(k_e)}$

Using momentum conservation, $q_{\parallel} = |\boldsymbol{k_e} - \boldsymbol{K}|$, we can express $R(q) = R(|\boldsymbol{k_e} - \boldsymbol{K}|)$. Taking the initial electron distribution to be $f_e(\boldsymbol{k_e})$, the total scattering rate into $\boldsymbol{K}$ is

$$G(\boldsymbol{K}) \propto \int d^2k_e\ f_e(\boldsymbol{k}_e)\, R(|\boldsymbol{k}_e - \boldsymbol{K}|).$$

which is essentially a convolution of $f_e(\boldsymbol{k_e})$ with $R(\boldsymbol{k_e})$. Since both $f_e$ and $R$ are decreasing and radially symmetric functions of $\boldsymbol{k_e}$, the function $G(\boldsymbol{K})$ is a monotonically decreasing function of $K$. (This can be easily seen by taking two radially symmetric functions and shifting them relative to each other – the overlap area (convolution) drops with the shift $K$). We conclude that the formation rate $G(K)$ is maximal at $K = 0$.

In practice, the IX population forms after a multi-phonon relaxation and exciton-exciton scattering a Bose distribution that peaks at $K = 0$. Once a small seed at $K = 0$ exists, Bose stimulation multiplies its feed by $(1 + N_0)$, so the zero-momentum mode wins the mode competition and clamps the reservoir, exactly the laser-like threshold mechanism.

## 3. Nuclear polarization rate equations

Let us assume the number of nuclei in an exciton volume is $N$, where $N_{\pm}$ are the number of up/down spins, such that $N_+ + N_- = N$, and define the nuclear polarization $P_N = \frac{(N_+ - N_-)}{N}$. Let us now take the spin flip rates up and down per nucleus as $W_{\pm}$, such that the total spin flip rates are $R_{\pm} = W_{\pm}N_{\mp}$. Each up/down-flip changes $P_N$ by $\pm\frac{2}{N}$, and therefore, the rate equation for $P_N$ can be written as

$$\frac{dP_N}{dt} = \frac{2}{N}(R_+ - R_-) - \frac{P_N}{T_N}$$

$$= (W_+ - W_-) - (W_+ + W_-)P_N - \Gamma_N P_N, \tag{4}$$

where $\Gamma_N$ is the (slow) nuclear polarization decay rate. The first term, which is due to the difference $\Delta W = W_+ - W_-$ in the up and down flip rates, can be viewed as a pump that tries to polarize the nuclei. The second term, which is proportional to the total flip rate, $\Sigma W = W_+ + W_-$ and is linear in $P_N$, is a backaction term, and together with the decay term they bring the system back into depolarized state.

Let us now write explicitly the rates $W_\pm$ using the Fermi's golden rule. If we do not include higher order recombination assisted process, we should consider the spectral overlap between the two exciton states. The spectral function that enters the golden-rule type expression is the Fourier transform of the excitonic correlation function,

$$S = \frac{\Gamma}{\Gamma^2 + \Delta^2}. \tag{5}$$

For a two-level system with pure radiative decays and no extra dephasing $\frac{\Gamma}{\hbar} = \frac{1}{T_2} = \frac{1}{2T_{1b}} + \frac{1}{2T_{1d}}$ where $T_{1b,d}$ is the decay rate of the bright (b) and dark (d) excitons. $\Delta$ is the energy separation between bright and dark and is given by $\Delta(P_N) = \Delta_0 - AP_N$, where $\Delta_0$ is the initial dark bright separation and $A$ is the Overhauser shift coefficient. So, the effective rates for the b → d and d → b channels are:

$$W_- = g_m^2\, n_b(1 + n_d)\frac{\Gamma}{\Gamma^2 + [\Delta - \hbar\omega_N]^2}, \tag{6}$$

$$W_+ = g_m^2\, n_d(1 + n_b)\frac{\Gamma}{\Gamma^2 + [\Delta + \hbar\omega_N]^2}. \tag{7}$$

Here $n_b, n_d$ are the $K = 0$ occupation of the bright and dark exciton states, respectively. The factor $g_m^2$ appears from the matrix element squared in the golden rule, $\hbar\omega_N$ is the nuclear Zeeman splitting, and we allowed for stimulated effects for both b → d and d → b channels. In our case, we can neglect the stimulation d → b in $W_+$.

From Eq. (4), at steady state,

$$P_N = \frac{1}{\Gamma_N}\left((W_+ - W_-) - (W_+ + W_-)P_N\right) \tag{8}$$

The figure below shows a graphical solution of Eq. (4) at steady state, where we plot the left- and right-hand sides of Eq. (8). Circles (crosses) indicate stable (unstable) solutions for $P_N$ [3].

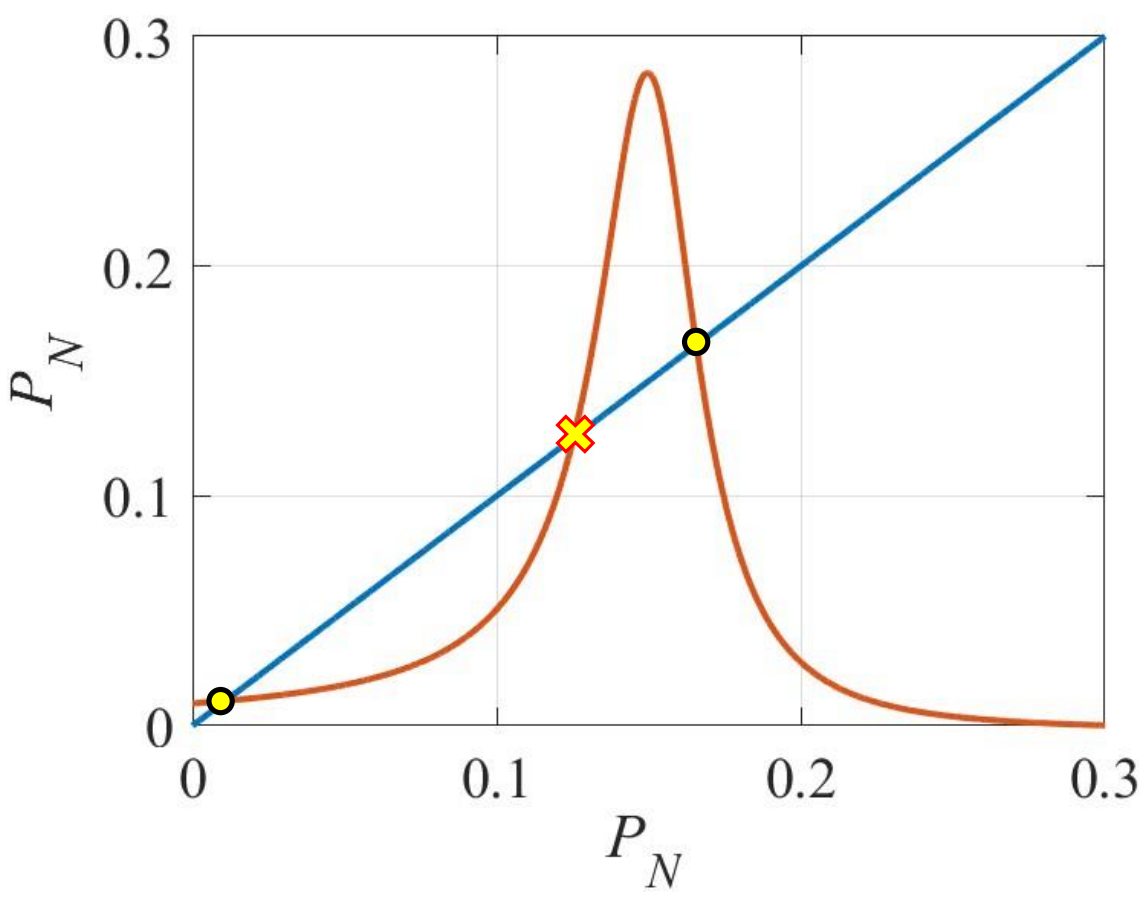

Fig. S4: A graphical solution of Eq. (8) showing bistability of $P_N$ using parameters $\frac{n_d}{n_b} = 15$, $\frac{\Delta_0}{A} = 0.15$, $g_m = 0.02$, $\Gamma_N = 1$, and $\Gamma = 0.01$.

Note that $W_\pm$ depend on the populations $n_b, n_d$ and these populations depend on the detuning $\Delta$, therefore, the full solution should be obtained by solving self-consistently the nuclear polarization rate equations together with the Lindblad master equations for the bright and dark excitons. However, a few insights can be obtained by examining the behavior in some limits:

## 4. RF radiation and condensate threshold

We start from the static Hamiltonian:

$$H_0 = \omega_e S_z + \omega_N I_z + \frac{a}{2}(S_+ I_- + S_- I_+) + a S_z I_z. \tag{9}$$

The flip-flop term only couples the two states $|\uparrow_e\downarrow_N\rangle$ and $|\downarrow_e\uparrow_N\rangle$. We define

$$\Delta \equiv \omega_e + \omega_N$$

In the basis $\{|\uparrow_e\downarrow_N\rangle, |\downarrow_e\uparrow_N\rangle\}$, the Hamiltonian takes the form (we take $|\omega_e + \omega_N| \gg a$)

$$H_0 = \frac{\Delta}{2}\,\sigma_z + \frac{a}{2}\,\sigma_x, \tag{10}$$

where $\sigma$ are Pauli matrices acting in that subspace. This can be diagonalized by a rotation with mixing angle $\theta$ defined by $\tan(2\theta) = \frac{a}{\Delta}$.

The eigenstates are

$$|+\rangle = +\cos\theta\ |\uparrow_e\downarrow_N\rangle\ + \sin\theta\ |\downarrow_e\uparrow_N\rangle,$$
$$|-\rangle = -\sin\theta\ \ |\uparrow_e\downarrow_N\rangle\ + \cos\theta\ \ |\downarrow_e\uparrow_N\rangle,$$

with splitting

$$\Omega_{\rm hf} = \sqrt{\Delta^2 + a^2}.$$

We now add RF modulation on the electron:

$$H_{\rm RF}(t) = \Omega_z \cos(\omega_{\rm RF} t)\, S_z. \qquad (11)$$

When we rotate into the eigen basis that diagonalizes $H_0$, the operator $\sigma_z$ becomes a combination of diagonal and off-diagonal terms:

$$\sigma_z \rightarrow \cos(2\theta)\ \tilde{\sigma}_z + \sin(2\theta)\ \tilde{\sigma}_x,$$

where $\tilde{\sigma}$ are Pauli matrices in the $\{|+\rangle, |-\rangle\}$ eigen-basis.

Therefore, the RF drive becomes

$$H_{\rm RF}(t) = \frac{\Omega_z}{2} \cos(\omega_{\rm RF} t)\, [\cos(2\theta)\ \tilde{\sigma}_z + \sin(2\theta)\ \tilde{\sigma}_x].$$

The RF drive has a longitudinal component, $\propto \tilde{\sigma}_z$, that just modulates the splitting, and a transversal component, $\propto \tilde{\sigma}_x$, that can induce transitions between the dressed $|+\rangle, |-\rangle$ eigenstates.

Now we go into a rotating frame at $\omega_{\rm RF}$ for the dressed two-level system and apply the rotating frame approximation. The resonance condition becomes simply:

$$\omega_{\rm RF} \approx \Omega_{\rm hf} = \sqrt{(\omega_e + \omega_N)^2 + a^2}$$

and at $|\omega_e + \omega_N| \gg a$, the resonance is essentially $\omega_e + \omega_N$, with a small hyperfine shift.

The important role of the RF is to help keeping the system at zero bright-dark detuning. So, we wish to protect our operating point against slow fluctuation due to the nuclear field, $\delta\Delta(\mathrm{t}) = A\delta I_z(t)$ that would shift the condensate from zero bright-dark detuning. A fluctuation $\delta\Delta(t)$ enters the Hamiltonian as

$$\delta H = \frac{\delta\Delta(t)}{2}\sigma_z.$$

If we rotate the fluctuation into the dressed eigen-basis, $\sigma_z \longrightarrow \tilde{\sigma}_x$. So it has a transversal component in the dressed basis, along the same axis as the RF:

$$\delta H_T = \frac{\delta\Delta(t)}{2}\,\tilde{\sigma}_x \tag{12}$$

The RF field couples such fluctuation to the electron's spin and allows damping of them through the fast decay of the electron polarization. That means that detuning noise is not stored for the characteristic long time of the nuclear spin decay: The injected energy, $\delta\Delta$, creates electron excitation, which is effectively dissipated into the bath.

## 5. The Overhauser field by the unpolarized nuclei

With depolarized nuclei (zero average polarization), the Overhauser field seen by an exciton is a random, zero-mean effective magnetic field arising from hyperfine coupling to many nuclei. Because the exciton samples a large number of nuclear spins, the field is very well-approximated by a 3D normal distribution (central limit theorem), with a width set by the effective number of nuclei within the exciton envelope.

Let $\mathbf{B}_N$ be the Overhauser field acting on the electron spin. For an unpolarized nuclear bath, $\langle \mathbf{B}_N \rangle = 0$, and the probability density for the vector field is isotropic Gaussian [4]:

$$P(\mathbf{B}_N) = \frac{1}{\left(2\pi\sigma_B^2\right)^{\frac{3}{2}}}\exp\left(-\frac{B_{N,x}^2 + B_{N,y}^2 + B_{N,z}^2}{2\sigma_B^2}\right), \tag{13}$$

with independent components

$$\left\langle B_{N,i}^2 \right\rangle = \sigma_B^2, \qquad i = x, y, z.$$

For contact hyperfine coupling, the Overhauser field variance can be written as

$$\sigma_B^2 = \frac{1}{3(g_e\mu_B)^2}\sum_\alpha a_\alpha^2\ I_\alpha(I_\alpha + 1)\ \frac{1}{N_{\mathrm{eff},\alpha}}$$

$$= \frac{5}{4(g_e\mu_B)^2}\sum_\alpha a_\alpha^2\ \frac{1}{N_{\mathrm{eff},\alpha}}\ ,$$

where $g_e$ is the electron $g$-factor, $\mu_B$ is the Bohr magneton, $\alpha$ runs over nuclear isotopes, $a_\alpha$ are the species-specific hyperfine constants, and we took $I_\alpha = \frac{3}{2}$.

$N_{\text{eff},\alpha}$ is the effective number of nuclei sampled by the carrier envelope

$$N_{\text{eff}} = \frac{1}{v_0 \int d^3r \, |\psi(\mathbf{r})|^4} \ ,$$

with $v_0$ the volume per nucleus (for zincblende GaAs, $v_0 \approx \frac{a_{\text{lat}}^3}{4}$ per atom/unit cell; $a_{\text{lat}}$ is the lattice constant).

We can calculate $N_{\text{eff}}$ assuming the total envelope wavefunction is $\psi(\mathbf{r}) = \phi(\boldsymbol{\rho})\chi(z)$, where $\phi(\boldsymbol{\rho})$ is the exciton is quasi-2D in-plane with a hydrogenic relative wavefunction

$$\phi(\boldsymbol{\rho}) = \frac{1}{\sqrt{2\pi a_B^2}} e^{-\frac{\rho}{a_B}},$$

and $\chi(z)$ is the carrier confinement in the ground QW sub band

$$\chi(z) = \sqrt{\frac{2}{L}} \cos\left(\frac{\pi z}{L}\right), \qquad z \in \left[-\frac{L}{2}, \frac{L}{2}\right]$$

For this choice one finds

$$N_{\text{eff}} = \frac{1}{v_0} \frac{64\pi a_B^2 L}{3}.$$

Then a common compact expression for $\sigma_B$ is

$$\sigma_B \sim \frac{a_{\text{rms}}}{|g_e|\mu_B} \sqrt{\frac{5}{4N_{\text{eff}}}} \approx \frac{a_{\text{rms}}}{16\,|g_e|\mu_B} \sqrt{\frac{15 v_0}{\pi a_B^2 L}},$$

where $a_{\text{rms}}$ is the abundance-weighted RMS hyperfine constant.

The variance of the Overhauser shift energy, $\sigma_N$, is then

$$\sigma_N \sim \frac{a_{\text{rms}}}{16} \sqrt{\frac{15 v_0}{\pi a_B^2 L}}.$$

Using $a_{\text{lat}} = 0.56$ nm, $a_B = 20$ nm, $L = 12$ nm, $a_{\text{rms}} \approx 50$ μeV [4] we get $\sigma_N \approx 5$ MHz.

Let us consider now a small magnetic field, ε, that opens a Zeeman gap between the $|+2\rangle$ and $|+1\rangle$ excitons states. As the nuclei become polarized and the Overhauser field, $B_N$, grows in magnitude opposite to the applied field, the total field acting on the electron spin, $B - B_N$, decreases, and the Zeeman gap closes. When $B_N$ exceeds the magnitude of $B$, the energy of the bright $|+1\rangle$ exciton becomes lower than that of the dark $|+2\rangle$ exciton, and the condensate becomes unstable. Therefore, the only stable realizations of nuclear polarization are those which satisfy $B_N \leq B$, and we are left with an asymmetric distribution of $B_N$, depicted by the black dashed line in the figure below. The mean value of the new distribution, $\langle B_N \rangle$, is shifted from $B$ to a lower field by $\Delta B_N$, given by

$$\Delta B_N = \frac{\int_{-\infty}^{\varepsilon} B_N \exp\left(-\frac{B_N^2}{\sigma_N^2}\right) dB_N}{\int_{-\infty}^{\varepsilon} \exp\left(-\frac{B_N^2}{\sigma_N^2}\right) dB_N}.$$

At the limit of small ε we get $\widetilde{\sigma}_N = -\frac{\sigma_N}{\sqrt{\pi}} \approx -2.8$ MHz, in excellent agreement with our experiment.

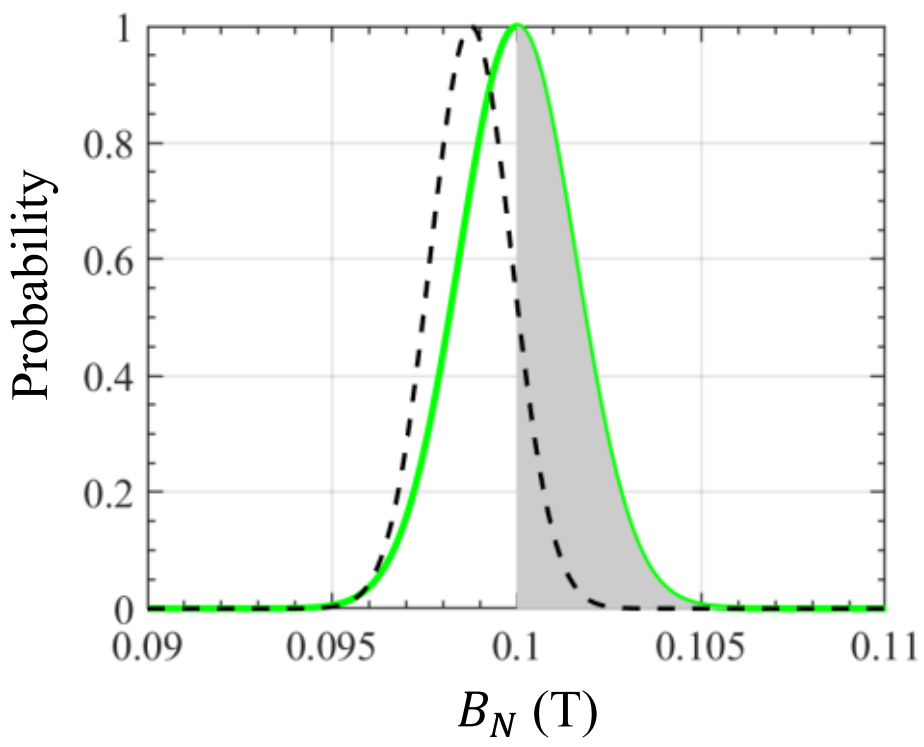

Fig. S5: The probability distribution for obtaining a certain Overhauser field, $B_N$, in an exciton volume, for $B = 0.1$ T. The grey area marks the probability to obtain $B_N > B$, and the black dashed line represents the allowed probability distribution, where $B_N \leq B$.

## 6. Extended data

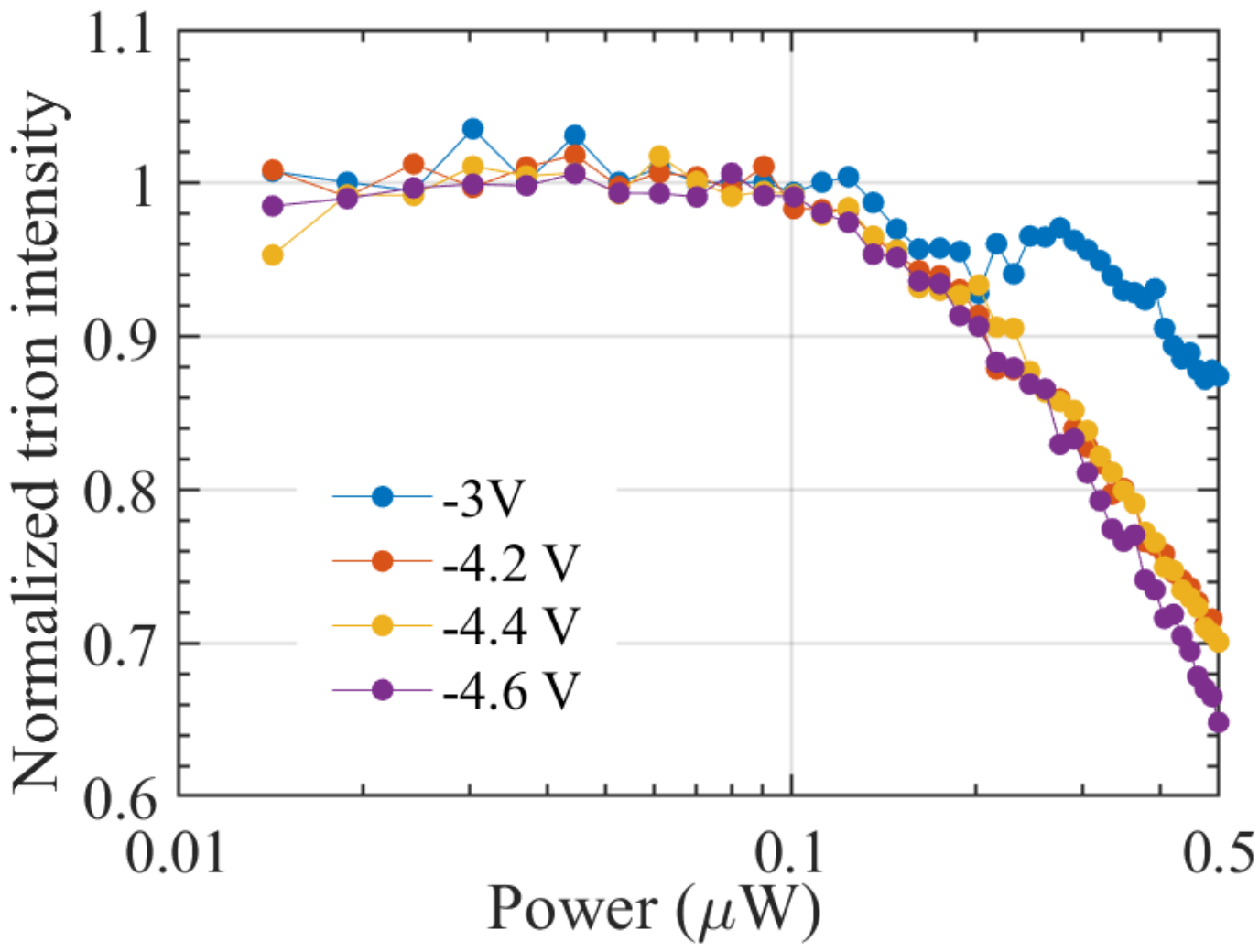

Fig. S6: The normalized trion intensity as a function of laser power for different gate voltages at $T = 1.7$ K. The trion intensities are first normalized by the total excitation power and subsequently scaled to their respective low-power values. Varying the gate voltage modifies the lifetime of bright IX and therefore their steady state density. However, the threshold power remains unchanged across all gate voltages. This establishes the fact that dynamics are governed solely by the dark IX band only.

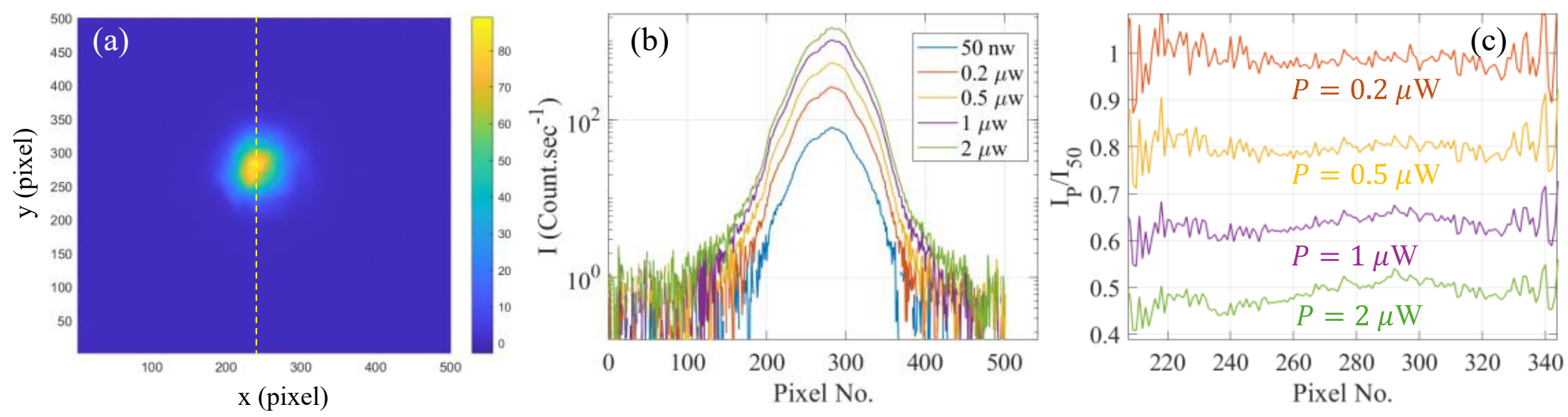

Fig. S7: (a) Raw PL image at an incident laser power of 50 nW. (b) Line cuts of the PL intensity taken along the dashed line in (a), shown for different incident laser powers ranging from 50 nW and 2 μW. Each pixel is 0.36 μm. (c) The reduction in normalized PL intensity with increasing power. Normalization is performed by dividing the curves of (b) by the corresponding total power and then by the sub-threshold cross-section at 50 nW. It is seen that darkening occurs uniformly across the excitation region, excluding the roll of local defects.

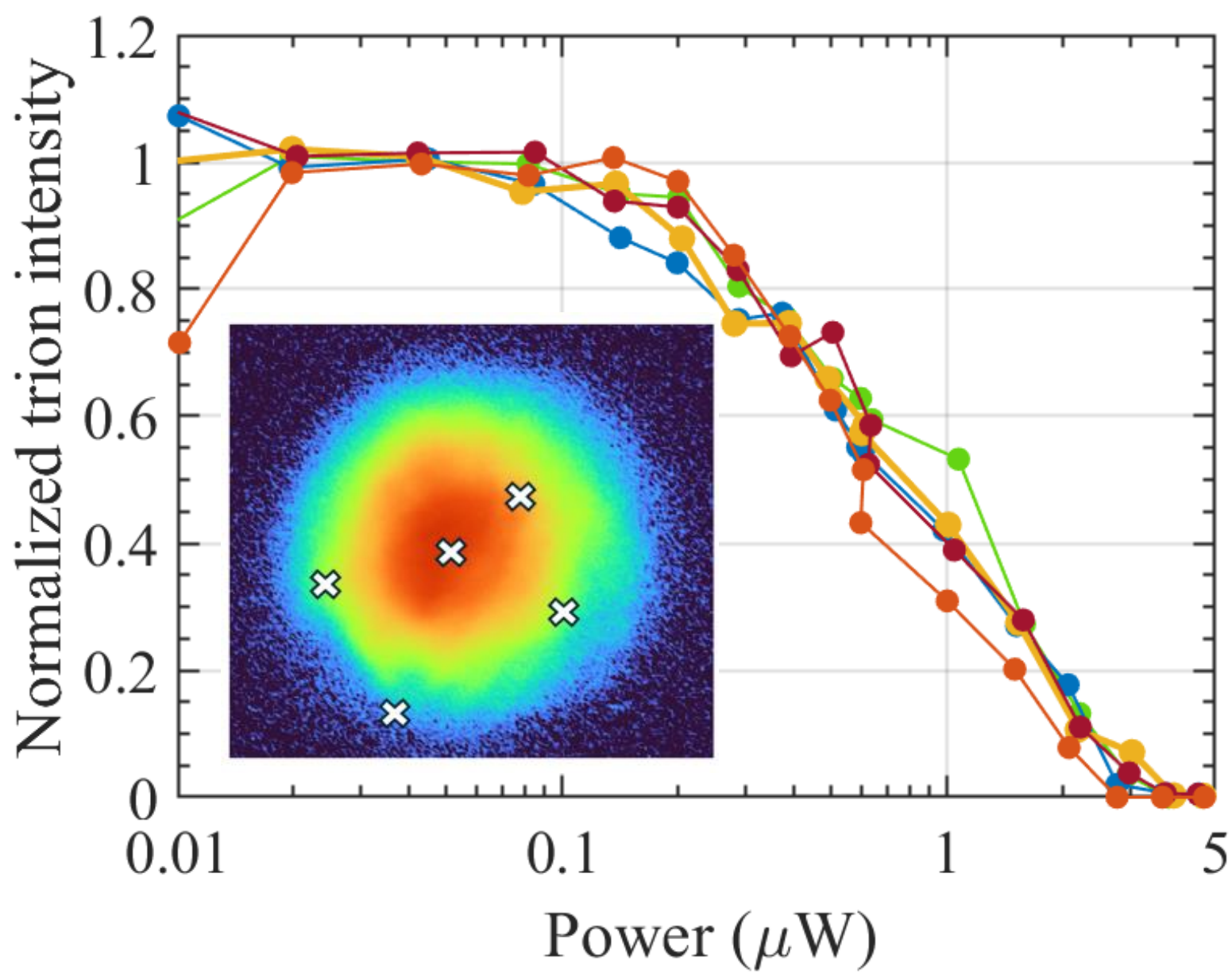

Fig. S8: Normalized trion intensity as a function of excitation power measured at different positions within the laser illumination spot. The trion intensities are scaled to their respective low-power values. The PL is collected at the crossed positions on the illuminated spot using a 20 µm pinhole at the image plane, corresponding to sub-micron at the sample plane. The consistent threshold behavior across all positions indicates that threshold is independent of local excitation power and occurs across the entire region.

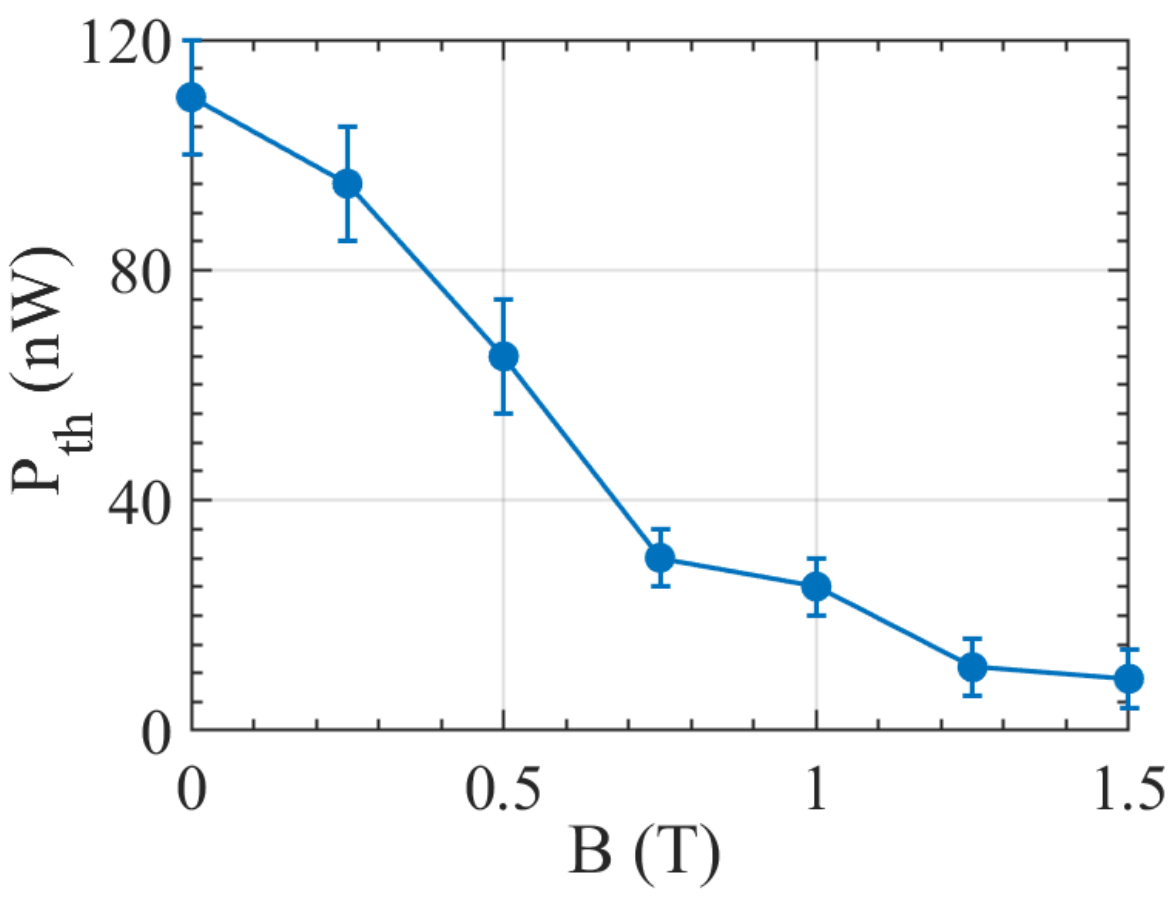

Fig. S9. Magnetic-field dependence of the threshold power $P_{\mathrm{th}}$. The threshold power is strongly affected by a magnetic field applied normal to the layers. As shown in figure, $P_{\mathrm{th}}$ decreases monotonically with increasing $B$, reaching only a few nW at $B = 1.7$ T. This trend is consistent with the expected suppression of heavy-hole/light-hole mixing and of the dark-bright exciton spin-flip rate in magnetic field, both of which reduce the condensate loss rate $\gamma_0$. If these channels dominate the losses, a decrease of the threshold power with magnetic field is naturally expected.

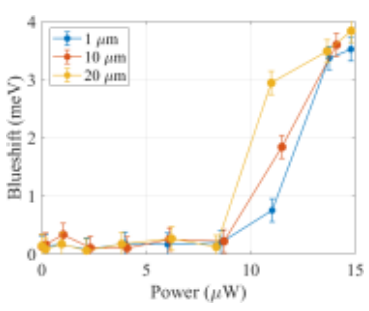

Fig. S10: Dependence of the blueshift on laser power for different excitation spot sizes at T=1.7 K. Remarkably, the onset power for the density buildup, $P_{up}$ is approximately the same for spot diameters ranging from 1 to 20 µm, although this range corresponds to a variation in local power density exceeding two orders of magnitude. This shows that the threshold is governed by the total injected power rather than by the local power density. The observation is consistent with spatial spreading of the condensate away from the excitation spot, such that the critical density is established over an extended area.

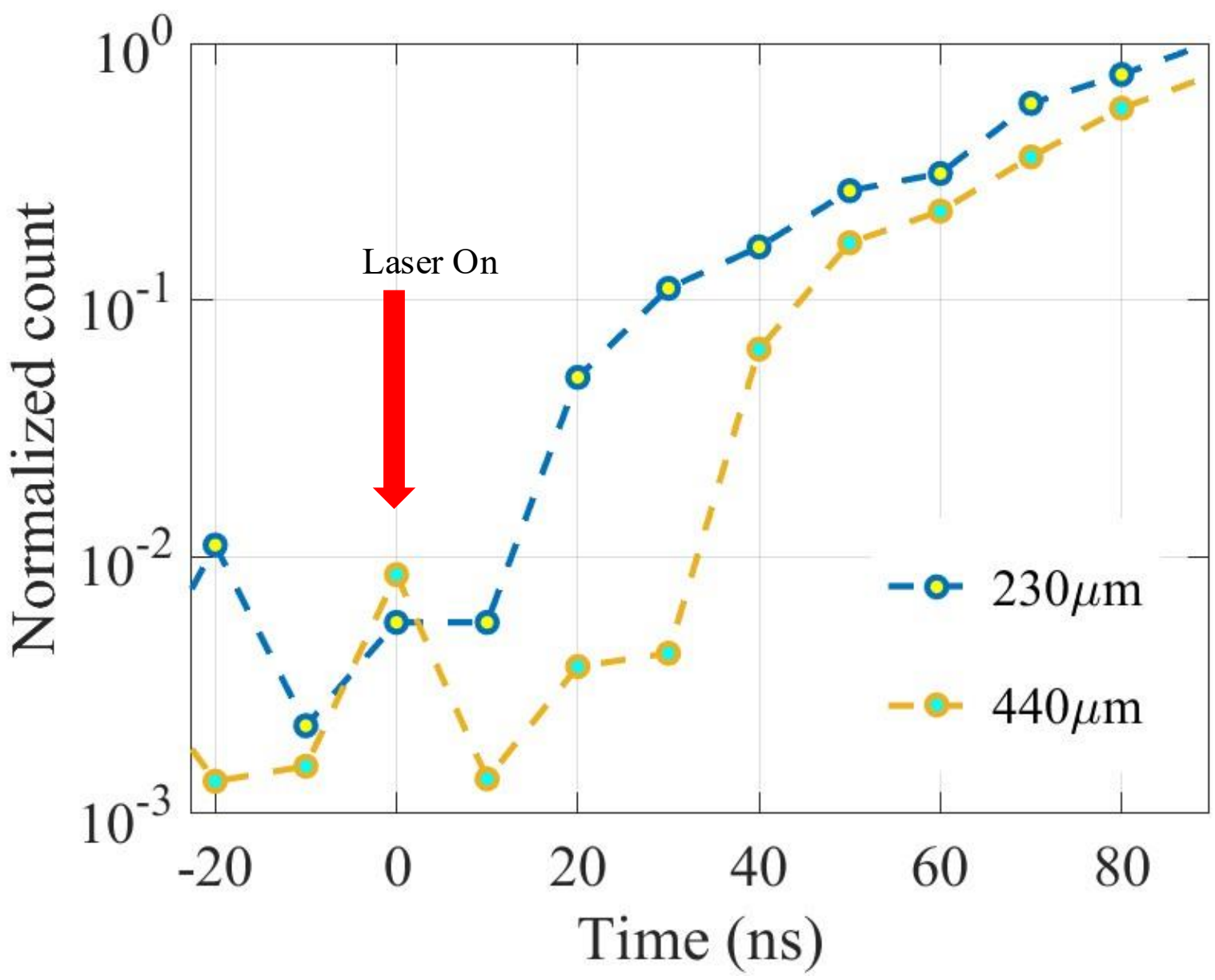

Fig. S11: The normalized photon counts at the different probe beam positions (see Fig. 2f in the main manuscript). At time = 0, the pump laser is switched on and persists for 2 μs. Additional check against optical crosstalk arises from the detection scheme. All signals are spectrally filtered through the spectrometer, which efficiently suppresses any residual laser stray light. Furthermore, the temporal reference (time $= 0$) is defined from the rising edge of the laser signal directly at the pump location, while the delay at the probe position is measured relative to this reference. This procedure ensures that no artificial propagation delays are introduced. Here $T = 0.6$ K.

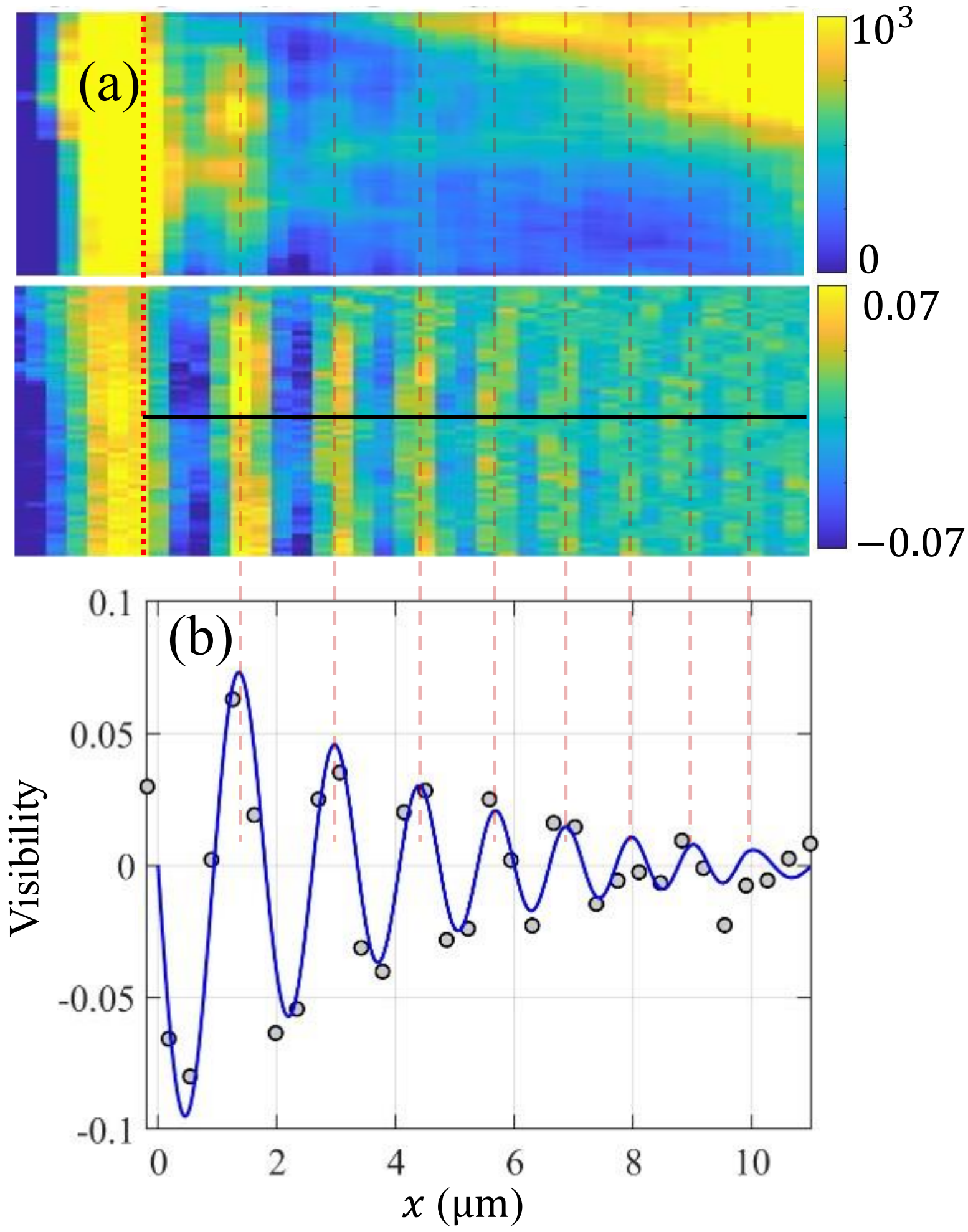

Fig. S12: a. PL image acquired close to the sample edge, where pronounced spatial oscillations are observed. Top panel shows the raw PL intensity, $I$, and bottom panel shows the visibility, $(I-\bar{I})/\bar{I}$ . Each pixel corresponds to 0.36 μm, and the red dotted line indicates the position of the top-gate edge. b. Visibility curve of the interference fringes extracted along the black solid line in Fig. a. A similar line-cut procedure is used for Figs. 3b-c in the MS. The blue solid line shows a fit to $e^{-x/l}\sin(2kx)$, with a $k = k_0 + \alpha x$. The fit yields $k_0 = 1.5 \pm 0.1\ \mu m^{-1}$, $\alpha = 0.09 \pm 0.01\ \mu m^{-2}$ and $l = 3 \pm 0.5\ \mu m$.

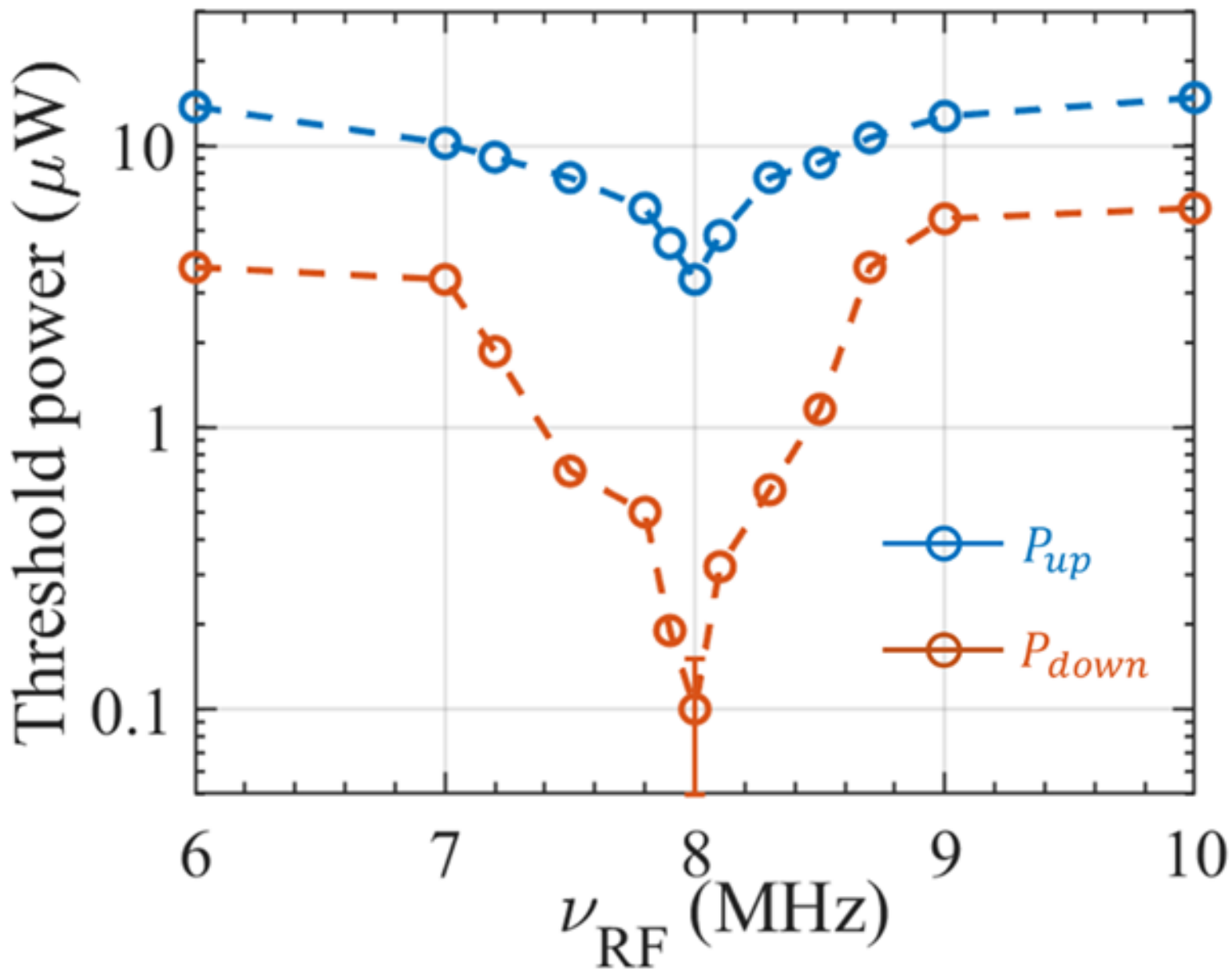

Fig. S13: Threshold power as a function of $\nu_{RF}$ at 1 T. Blue (orange) circles correspond to ramping up (down) the power. This shows that Overhauser field is fixed robustly locked to external field. This unique property of the condensate-nuclear system is nicely demonstrated here, which shows that the RF resonance, which is obtained when ramping the power up or down, is at the same frequency even though the optical power differs by almost two orders of magnitude. The resonance behavior is shown for $^{75}$As.

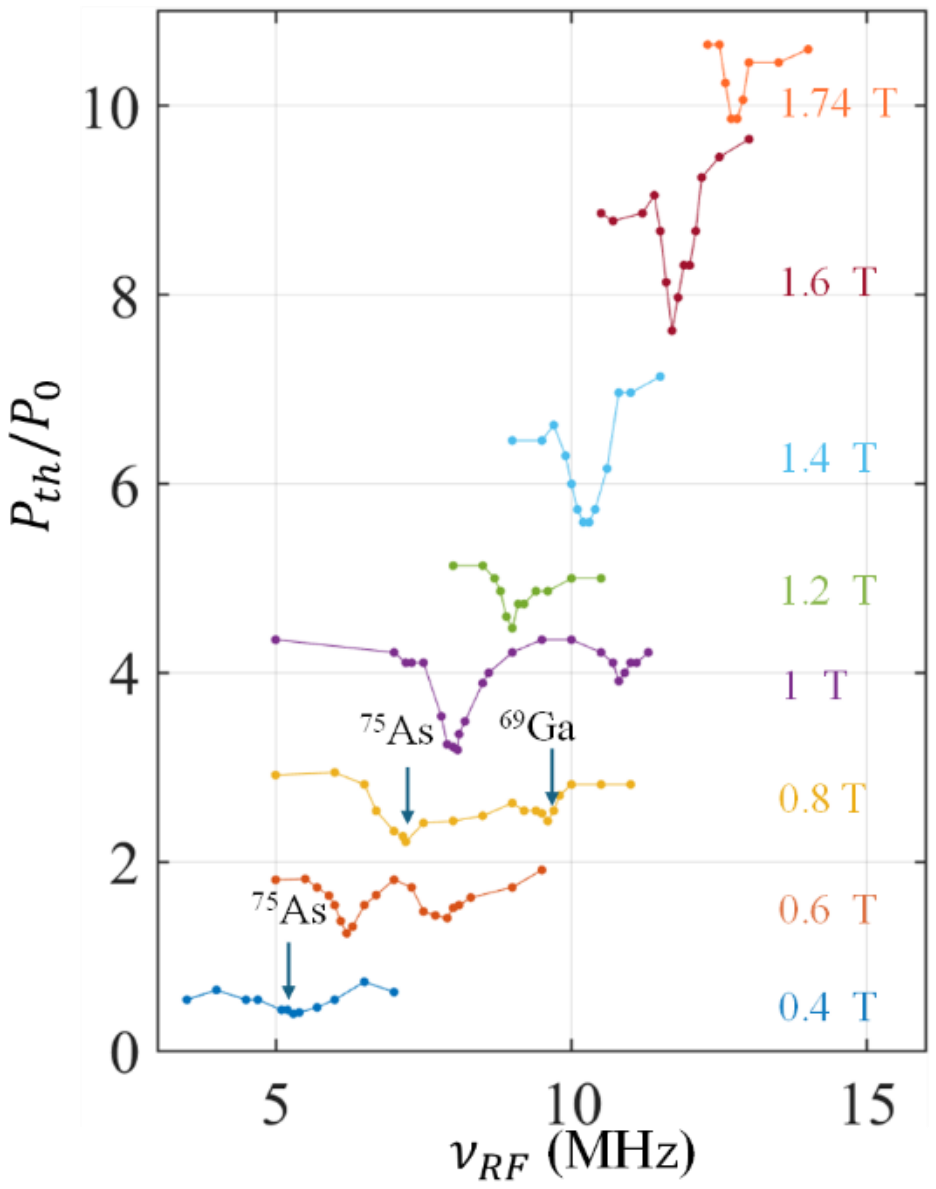

Fig. S14: shows $P_{th}$ as a function of $\nu_{RF}$ at various magnetic fields. We have scaled $P_{th}$ at each magnetic field by $P_0$, the value of $P_{th}$ at $\nu_{RF} = 0$, and shifted the curves vertically for better visualization. The lower frequency peak seen in all curves is the $^{75}$As peak, while the higher frequency one seen at 0.6 T, 0.8 T, and 1 T curves, corresponds to $^{69}$Ga. It is clearly seen that the resonance frequencies of both $^{75}$As and $^{69}$Ga increase monotonically with the magnetic field, as depicted in Fig. 5b of the main manuscript. We note that the resonance is broad at low magnetic fields, ~ 1 MHz at 0.4 T, and narrows as the field increases, 0.3 MHz at 1.6 T. This narrowing reflects the diminishing of the fluctuations of the Overhauser field due to the unpolarized nuclei with increasing field strength.

# 7. References:

1. A. Jash, M. Stern, S. Misra, V. Umansky, I. B. Joseph, Giant hyperfine interaction between a dark exciton condensate and nuclei. *Sci. Adv.* **10**, eado8763 (2024).
2. I. Carusotto, C. Ciuti, Quantum fluids of light. *Rev. Mod. Phys*. **85**, 299–366 (2013).
3. P. Maletinsky, C. W. Lai, A. Badolato, A. Imamoglu, Nonlinear dynamics of quantum dot nuclear spin. *Phys. Rev. B* **75**, 035409 (2007).
4. I. A. Merkulov, Al. L. Efros, M. Rosen, Electron spin relaxation by nuclei in semiconductor quantum dots. *Phys. Rev. B* **65**, 205309 (2002).